\documentclass{article} %
\usepackage[T1]{fontenc}
\usepackage{algorithmicx}
\usepackage{colm2024_conference}

\usepackage{booktabs}
\usepackage{graphicx}
\usepackage{enumitem}
\usepackage{wrapfig}
\usepackage{algorithm}
\usepackage{algpseudocode}
\usepackage{multicol}

\usepackage{microtype}
\usepackage{amsmath}
\usepackage{amsfonts}
\usepackage{colortbl}
\usepackage[utf8]{inputenc}
\definecolor{lightgray}{rgb}{0.9,0.9,0.9}
\usepackage{caption}
\usepackage{subcaption}
\usepackage{xcolor}
\usepackage{setspace}
\usepackage{url}
\usepackage{multirow}
\usepackage{colortbl}
\usepackage{tabularx}
\usepackage{xltabular}
\usepackage{blindtext}
\usepackage{pgfplots}
\pgfplotsset{compat=1.18}
\usepackage{tikz}
\usetikzlibrary{er, positioning, bayesnet}
\usepackage{makecell}
\usepackage{tipa}
\usepackage{siunitx}
\usepackage{nicefrac}
\usepackage{tocloft}
\usepackage{listings}
\usepackage[raster, skins]{tcolorbox} %
\usepackage{xltabular}
\usepackage{adjustbox}
\usepackage{xurl}
\usepackage{multirow}
\usepackage{threeparttable}
\usepackage{xcolor}
\usepackage{soul}

\usepackage{amsmath,amssymb,amsthm}
\usepackage{mathtools}

\usepackage{amsmath,amsfonts,bm}

\def\eqref#1{equation~\ref{#1}}

\def\1{\bm{1}}

\DeclareMathAlphabet{\mathsfit}{\encodingdefault}{\sfdefault}{m}{sl}
\SetMathAlphabet{\mathsfit}{bold}{\encodingdefault}{\sfdefault}{bx}{n}

\title{Qwen-Music Technical Report}

\author{ \bf Qwen Team }

\usepackage{pifont}
\newcommand{\cmark}{\ding{51}}
\newcommand{\xmark}{\ding{55}}

\begin{document}
\maketitle

\begin{abstract}
In this report, we introduce Qwen-Music, a powerful music generation model capable of producing highly musical and high-fidelity songs with complete vocal singing.
Qwen-Music supports two core tasks: \textbf{Text to Music Generation}, which create entirely new songs from text descriptions, lyrics, and musical attributes, and \textbf{Cover Song Generation}, which reinterprets existing songs with different styles and vocal characteristics.
Architecturally, Qwen-Music integrates three core components: \textbf{Qwen-Music-Tokenizer}, \textbf{Qwen-Music-LLM}, and \textbf{Qwen-Music-Render}.
Qwen-Music-Tokenizer compresses audio into a 25 Hz single-codebook stream of Music Semantic Tokens that preserve semantic and melodic information for LLM prediction.
Based on these tokens, Qwen-Music-LLM performs autoregressive music semantic modeling, with a key novelty being a melody-token-based chain-of-thought (Melody-CoT) mechanism that plans melodies before full-song generation, improving creativity, musicality, structural coherence, and reference-audio-based melody cloning.
To overcome the fidelity limitations of discrete semantic tokens, Qwen-Music-Render performs generative stereo rendering, enriching acoustic details and producing high-fidelity stereo waveforms.
We first apply a quality-aware pre-training curriculum to Qwen-Music-LLM, followed by progressive post-training comprising supervised initialization, offline DPO, and online GSPO, to further improve musicality and instruction-following ability.
Using 600 evaluation inputs, each comprising AI-generated lyrics and musical tags, Qwen-Music achieves state-of-the-art results in 13 of 16 objective musicality and audio-quality metrics.
Professional evaluators also prefer Qwen-Music over leading proprietary systems. 
For cover song generation, Qwen-Music preserves reference melodies more accurately than Suno V5.5, Suno V5, and MiniMax Cover on the AI-generated reference set, and outperforms MiniMax Cover on most metrics in the real-world popular-song reference set.
\end{abstract}
\begin{figure}[tbh]
\centering
\includegraphics[width=0.9\textwidth]{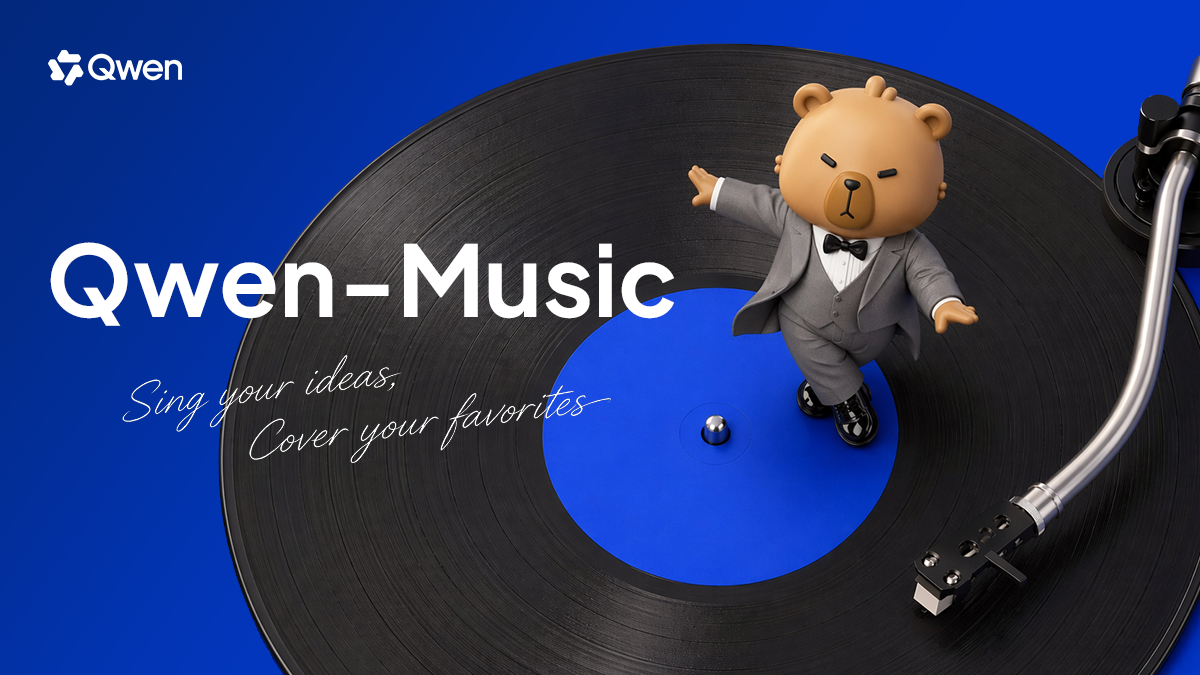}
\caption{Qwen-Music is a powerful and controllable music generation model capable of producing songs with complete vocal singing. It generates complete songs from text descriptions, lyrics, and musical attributes, and uses reference audio for cover song generation with different styles and vocal characteristics.}
\label{fig:intro}
\end{figure}
\section{Introduction}
\label{sec:intro}
Music generation has recently emerged as a key challenge in generative modeling, aiming to synthesize music that is musically coherent, acoustically natural, and semantically aligned with user intent \citep{agostinelli2023musiclm, mousai, musicgen, melody, majumder2024tango, noise2music, chen2024musicldm}.
Unlike general audio generation, song generation requires the joint modeling of lyrics, melody, rhythm, vocal performance, instrumentation, and musical structure over minutes \citep{yang2026heartmula, lei2026levo2stablemelodious, lei2026levo, liu2025songgen, gong2025ace, gong2026ace, yuan2025yue, ning2025diffrhythm, jiang2025diffrhythm, lei2024songcreator}.
A useful system should support open-ended song generation from text descriptions and lyrics while allowing users to control musical attributes such as genre, mood, instrumentation, and vocal timbre.
It should also reinterpret existing songs by preserving a reference melody while changing style or vocal characteristics.
Despite rapid progress in large-scale audio generation models, generating songs with stable melodic development, clear lyric articulation, realistic singing voices, and natural instrumental accompaniment remains challenging.

The central technical challenge is the mismatch between semantic composition and acoustic rendering. At the semantic level, a model must plan lyrics, vocal melody, section structure, repetition, and stylistic progression over long horizons. At the acoustic level, it must render timbre, phase, stereo placement, and high-frequency detail at waveform resolution. Discrete-token language models provide a scalable interface for global sequence generation, but compressed tokens lose acoustic detail \citep{xu2025mucodec, Encodec, yang2026heartmula}. Reference-guided cover song generation adds another constraint: the system must reuse the melody while allowing controllable changes to the arrangement, singing voice, and style.

In this report, we introduce \textbf{Qwen-Music}, a large-scale music generation system that separates song generation into semantic composition and acoustic rendering.
Qwen-Music supports both text-to-music generation and reference-audio-based cover song generation within a single framework.
Given text descriptions, lyrics, musical attributes, and optional reference audio, Qwen-Music can generate complete songs with controllable genre, mood, instrumentation, vocal timbre, and vocal gender.
Qwen-Music generalizes strongly across diverse languages, genres, and musical styles.

\begin{figure}[h]
  \centering
  \includegraphics[width=0.8\linewidth]{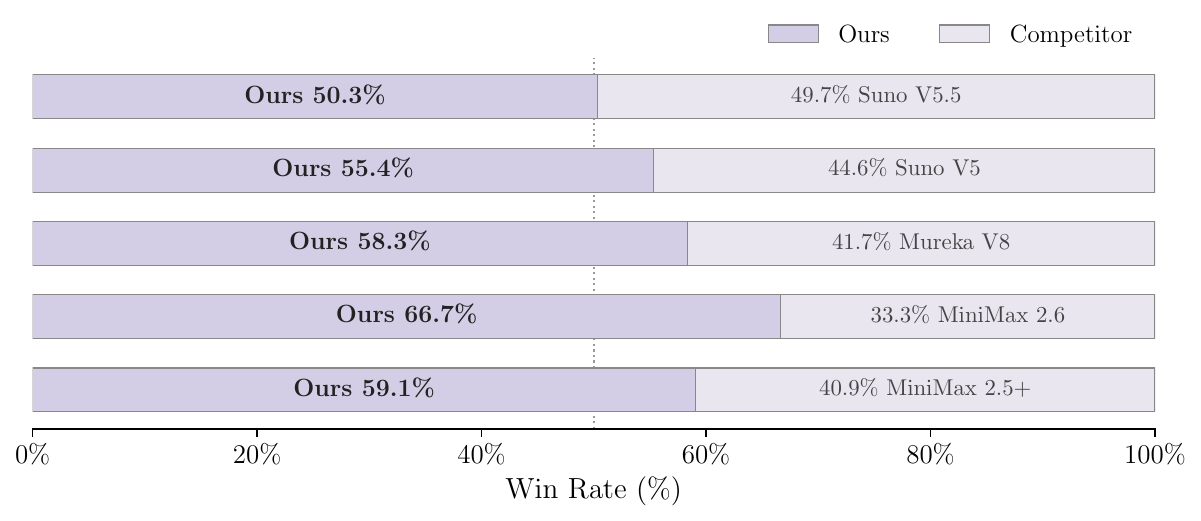}
  \caption{Blind A/B preference results between Qwen-Music and leading proprietary systems, including MiniMax Music 2.5+, MiniMax Music 2.6, Mureka V8, Suno V5, and Suno V5.5, judged by professional human raters. Each pair is evaluated under anonymized and randomized presentation, and values indicate the percentage of expert votes preferring each system.}
  \label{fig:ab_preference}
\end{figure}

Qwen-Music is designed around four key principles:

\begin{itemize}
\item \textbf{Composition in a compact semantic music space.} Instead of directly modeling waveforms, Qwen-Music represents music as 25 Hz single-codebook Music Semantic Tokens. This compact stream gives the autoregressive model a tractable sequence space for full-song composition while preserving the musical semantics needed by the renderer.

\item \textbf{Explicit melody planning and control.} To bridge the gap between high-level textual intent and concrete musical realization, Qwen-Music-LLM makes melody an explicit intermediate representation. For text-to-music generation, Melody-CoT plans the vocal melody before full-token generation. For cover generation, reference melody tokens condition the sequence so the output follows the source melody under the new style and vocal attributes.

\item \textbf{Generative acoustic rendering.} Qwen-Music-Render treats Music Semantic Tokens as musical sketches for generative rendering. A semantic-conditioned DiT predicts acoustic latents, the Spec-VAE decoder reconstructs spectrograms, and the Band-Mode Refiner corrects frequency-dependent magnitude and phase detail before waveform synthesis.

\item \textbf{Quality-graded scaling and multi-stage preference alignment.}
To effectively exploit large-scale heterogeneous music corpora, Qwen-Music organizes pre-training around a quality-graded data curriculum, progressively learning from data at different quality levels.
It further adopts a multi-stage post-training pipeline that combines supervised learning, offline preference optimization, and online policy optimization to improve musicality, instruction following, and controllability.
\end{itemize}

\begin{figure}[t]
  \centering
  \includegraphics[width=0.7\linewidth]{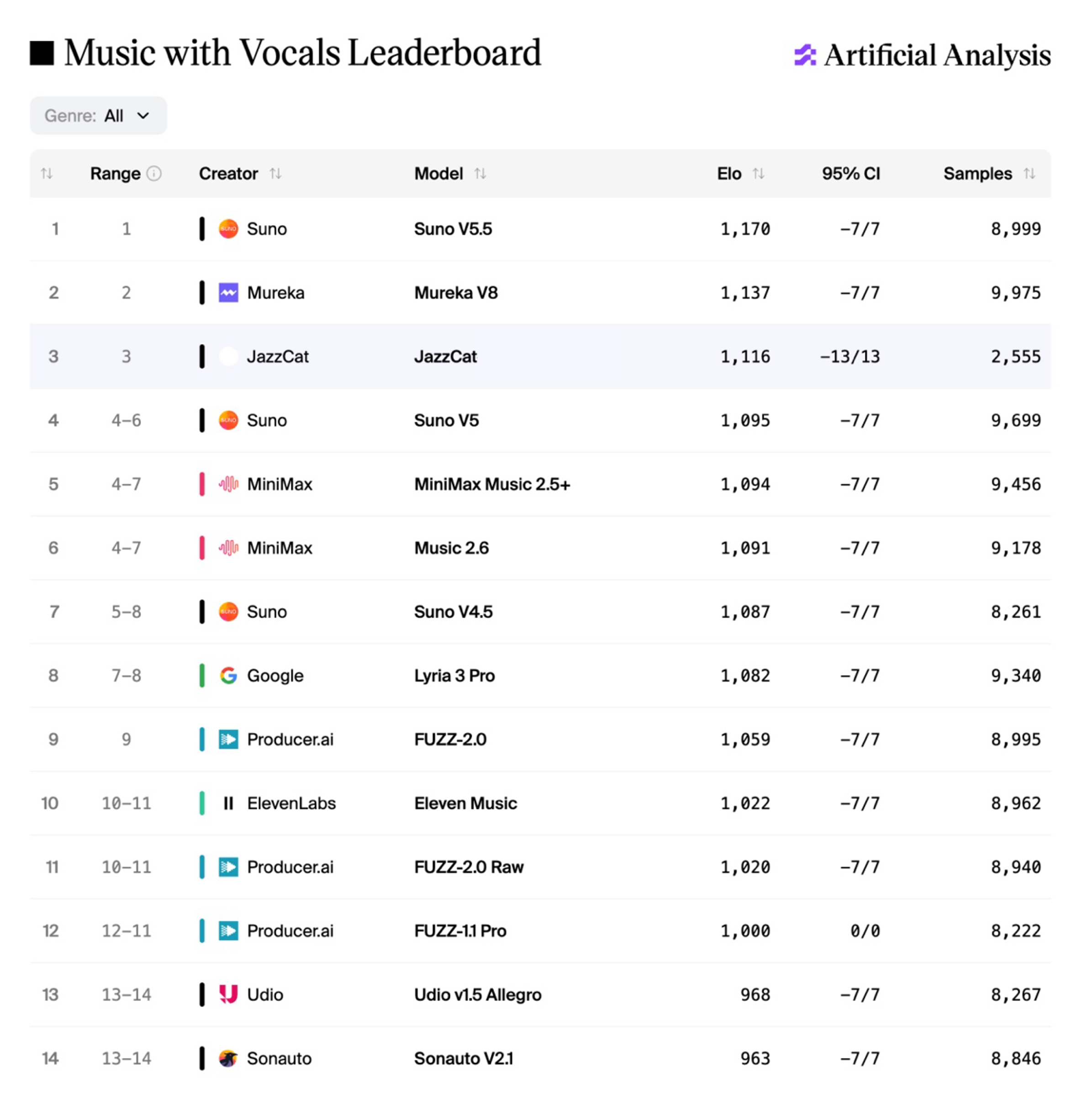}
  \caption{External leaderboard result on the Artificial Analysis Music with Vocals Leaderboard. Qwen-Music participated as \texttt{JazzCat} and ranked among the leading English vocal music generation systems.}
  \label{fig:jazzcat_aa}
\end{figure}

Empirically, Qwen-Music demonstrates strong competitiveness against leading proprietary music generation systems.
As shown in Figure~\ref{fig:ab_preference}, blind A/B preference tests judged by professional human raters show that Qwen-Music achieves win rates of 59.1\% against MiniMax Music 2.5+, 66.7\% against MiniMax Music 2.6, 58.3\% against Mureka V8, and 55.4\% against Suno V5, while remaining comparable to Suno V5.5 with a slight preference advantage of 50.3\%.
Beyond our internal evaluation, Qwen-Music also appears as \texttt{JazzCat} on the Artificial Analysis Music with Vocals Leaderboard\footnote{\url{https://artificialanalysis.ai/music/leaderboard/vocals}}, ranking third among leading English vocal music generation systems (Figure~\ref{fig:jazzcat_aa}).
Objective evaluation in Section~\ref{sec:experiment} further shows that Qwen-Music obtains the best result on 13 out of 16 text-to-music metrics across SongBench \citep{wu2026songbench}, SongEval \citep{yao2025songeval}, and AudioBox-Aesthetic \citep{tjandra2025meta}, while cover song generation results demonstrate strong reference-melody preservation, stylistic adaptation, and vocal controllability.

Overall, Qwen-Music provides a unified framework for controllable, high-fidelity, and structurally coherent music generation. It bridges semantic-level composition and waveform-level synthesis, enabling both open-ended song creation and reference-guided musical reinterpretation in a single, coherent system.

\section{Architecture}
\label{sec:architecture}
\begin{figure}[tbh]
    \centering
    \includegraphics[width=0.9\textwidth]{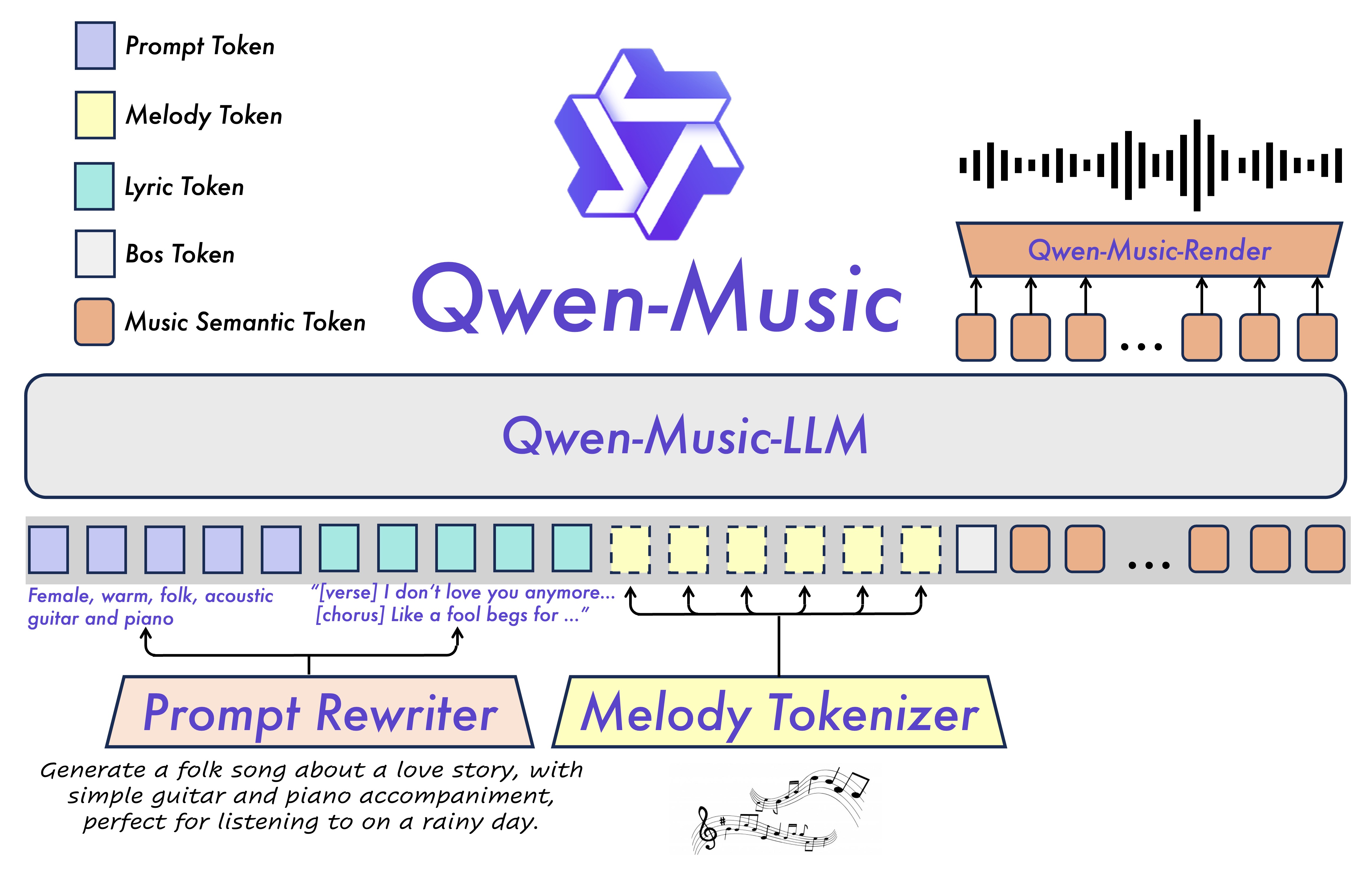}
    \caption{
      Overview of the Qwen-Music inference pipeline.
      Given a user request, Qwen-Music first rewrites the natural-language description into a structured textual condition, including musical tags (such as genre, singer characteristics, instrumental arrangement, etc.) and generated lyrics.
      Qwen-Music-LLM then generates Music Semantic Tokens, optionally using melody tokens extracted from a reference song for melody cloning.
      Finally, Qwen-Music-Render takes both the rewritten textual condition and the generated Music Semantic Tokens as input and performs generative rendering to produce high-fidelity 48 kHz stereo waveforms.
    }
    \label{fig:ovewview_arch}
\end{figure}

\subsection{Overview}


As shown in Figure~\ref{fig:ovewview_arch}, Qwen-Music follows an inference pipeline that transforms a user request into a complete high-fidelity stereo song.
Given a natural-language description, the system first rewrites it into a structured textual condition containing musical tags (such as genre, singer characteristics, instrumental arrangement, etc.) and generated lyrics.
Conditioned on this rewritten text, Qwen-Music-LLM autoregressively predicts Music Semantic Tokens, optionally incorporating reference melody tokens via Melody-CoT conditioning for cover song generation.
Qwen-Music-Render then takes both the rewritten textual condition and the generated Music Semantic Tokens as input and performs generative rendering to produce high-fidelity 48 kHz stereo waveforms.
This pipeline is supported by three core components: Qwen-Music-Tokenizer, Qwen-Music-LLM, and Qwen-Music-Render.

\begin{itemize}
\item \textbf{Qwen-Music-Tokenizer} maps raw music audio into 25 Hz Music Semantic Tokens using a Conformer-based encoder. It is trained with a four-stage recipe that combines self-supervised learning, causal adaptation, multi-task supervision, and vector quantization. The resulting token stream provides the discrete music representation used for language-model training and renderer conditioning.

\item \textbf{Qwen-Music-LLM} generates Music Semantic Tokens conditioned on text descriptions, lyrics, musical attributes, and optional melody tokens. For original song generation, it can first produce a Melody-CoT plan before generating the full Music Semantic Token sequence. For cover song generation, it uses melody tokens extracted from a reference song as conditioning context while following the target style and vocal attributes.

\item \textbf{Qwen-Music-Render} converts generated Music Semantic Tokens into 48 kHz stereo waveforms. It uses a semantic-conditioned Diffusion Transformer to predict continuous acoustic latents, a spectral-domain VAE decoder to reconstruct complex spectrograms, and a Band-Mode Refiner to correct frequency-dependent magnitude and phase details before inverse STFT synthesis.
\end{itemize}

In the following sections, we first introduce Qwen-Music-Tokenizer and its multi-stage training recipe. We then describe Qwen-Music-LLM, including melody tokenization and melody-aware generation. Finally, we present Qwen-Music-Render, focusing on diffusion-based latent generation, spectral-domain decoding, and band-wise refinement.

\subsection{Qwen-Music-Tokenizer}

\begin{figure}[tbh]
    \centering
    \includegraphics[width=1\textwidth]{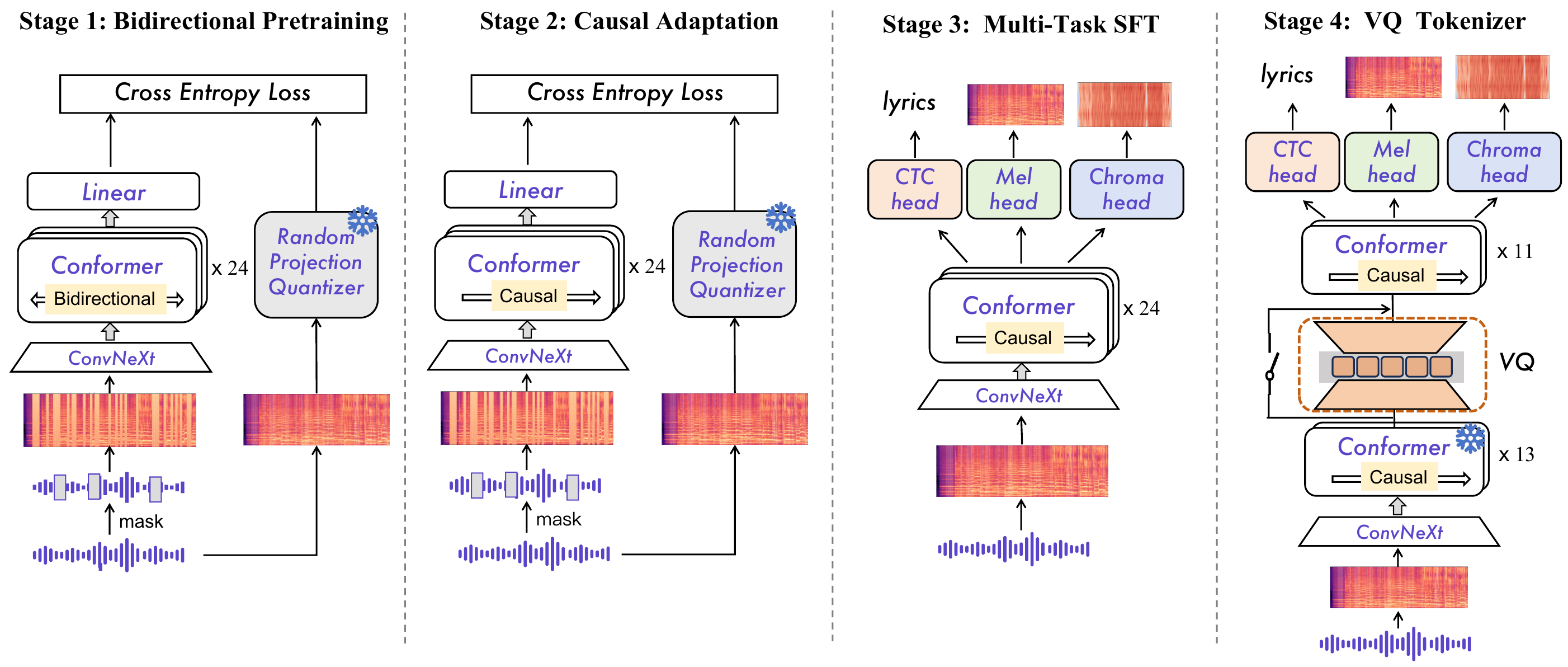}
    \caption{
        Overview of Qwen-Music-Tokenizer. The tokenizer maps music waveforms to 25 Hz Music Semantic Tokens through BestRQ pretraining, causal adaptation, multi-task SFT, and VQ tokenizer training.}
    \label{fig:tokenizer}
\end{figure}

Qwen-Music-Tokenizer is a low-bitrate discrete tokenizer that maps a
raw music waveform into a single stream of 25 Hz Music Semantic Tokens. It uses a single backbone consisting of a convolutional frontend followed by a Conformer encoder~\citep{conformer}. As illustrated in Figure~\ref{fig:tokenizer}, we
train the backbone in \emph{four} stages: (i)~bidirectional self-supervised pretraining with BestRQ~\citep{bestrq}, (ii)~causal adaptation of the pretrained encoder,
(iii)~multi-task supervised fine-tuning (SFT) with lyric and spectral targets,
and (iv)~VQ tokenizer training. The final stage inserts the VQ
bottleneck and trains the final tokenizer. Stages 2--4 are initialize from the
previous checkpoint, while Stage 1 starts from random initialization. This
recipe first learns continuous music representations and then introduces causal
attention, supervision, and vector quantization. The deployed tokenizer keeps
the causal encoder up to the VQ insertion point and produces the $25$~Hz token
stream consumed by downstream autoregressive models.
Table~\ref{tab:tok-recipe}
summarizes the recipe.

\subsubsection{Backbone Architecture}
\label{sec:tok-arch}

The tokenizer takes raw music audio as input and encodes it into a compact low-frame-rate token sequence.
Specifically, each audio clip is converted to a mono $24$~kHz waveform and represented using log-Mel spectrogram features.
A lightweight convolutional frontend reduces the frame rate to $25$~Hz, corresponding to one frame every $40$~ms.
The tokenizer backbone is a $24$-layer, $0.6$B-parameter Conformer~\citep{conformer}, which is shared across all four training stages.
Starting from Stage~2, the backbone is converted to causal self-attention to match the downstream autoregressive modeling setting.
When encoding audio, the tokenizer is applied offline to the full clip and produces a compact $25$~Hz token stream.

\subsubsection{Stage 1: Bidirectional BestRQ Pretraining}
\label{sec:tok-stage1}

We begin with a self-supervised pretraining stage to build a general-purpose music representation from unlabeled audio.
The encoder is trained \emph{bidirectionally} with BestRQ~\citep{bestrq}, a masked-prediction objective whose targets are produced by a \emph{frozen} random quantizer, removing the need for a separately learned target network.

\paragraph{Random-projection targets}
Target codes are formed from the clean log-Mel spectrogram: a local context
window is pooled with temporal stride $4$ so that targets align with the $25$~Hz
encoder frames, projected by a fixed random projection, and mapped to the nearest
entry of a fixed, $\ell_2$-normalized random codebook. The projection and
codebook are randomly initialized and kept frozen throughout training, so no
target network is learned.

\paragraph{Masking and objective}
We apply time-domain span masking: the waveform is partitioned into $0.4$~s
spans, each masked independently with probability $0.3$ (guaranteeing at least
one span per sample), and masked spans are replaced with low-variance Gaussian
noise. The encoder consumes the masked audio, and a lightweight prediction head
is trained to recover the frozen random-projection targets at masked frames under
a cross-entropy loss,
\begin{equation}
\mathcal{L}_{\mathrm{BestRQ}}
= \frac{1}{|\mathcal{M}|} \sum_{t \in \mathcal{M}}
\operatorname{CE}\bigl(p(h_t), q(x_t)\bigr).
\end{equation}
where $\mathcal{M}$ is the set of masked frames, $h_t$ is the encoder output at
frame $t$, $q(\cdot)$ the frozen random-projection quantizer, and $p(\cdot)$ the
prediction head.

\subsubsection{Stage 2: Causal Adaptation}
\label{sec:tok-stage2-causal}

To prevent the encoder from accessing future frames and to align its
representations with downstream autoregressive modeling, we restrict
self-attention to be causal. Rather than training a causal encoder from
scratch, we resume BestRQ pretraining from the Stage~1 checkpoint after
switching self-attention from full bidirectional masking to causal
(left-context-only) masking, while leaving the pretraining objective otherwise
unchanged. This short adaptation stage converts the encoder to causal
self-attention at a small fraction of the Stage~1 cost, while preserving much
of the representation learned under the less constrained bidirectional regime.
All subsequent stages inherit causal attention.

\subsubsection{Stage 3: Multi-Task Supervised Fine-Tuning}
\label{sec:tok-stage3-sft}


Before quantization, we use multi-task supervision to make the encoder representations more informative for both language and music.
In particular, this stage encourages the future discrete codes to preserve lyrical content together with musical semantic attributes such as melody and harmony.

We attach three lightweight heads to the $25$~Hz
encoder outputs: (i)~a linear \emph{CTC head}~\citep{ctc} that transcribes lyrics
over a multilingual subword vocabulary; (ii)~a ConvNeXt
\emph{Mel head} that upsamples the latents back to the $100$~Hz, $128$-bin Mel
spectrogram; and (iii)~a ConvNeXt \emph{chroma head} predicting the $12$-bin
chroma feature. The objective combines the CTC loss with spectral-reconstruction
losses on Mel and chroma, each a sum of a spectral-convergence term and a
magnitude term,
\begin{equation}
\mathcal{L}_{\text{SFT}}
= \lambda_{\text{ctc}}\,\mathcal{L}_{\text{CTC}}
+ \lambda_{\text{mel}}\,\mathcal{L}_{\text{spec}}^{\text{mel}}
+ \lambda_{\text{chr}}\,\mathcal{L}_{\text{spec}}^{\text{chroma}},
\qquad
\lambda_{\text{ctc}}\!=\!\lambda_{\text{mel}}\!=\!\lambda_{\text{chr}}\!=\!1 .
\end{equation}
Input masking is effectively disabled ($p_{\text{mask}}\!=\!0.01$) so the model
sees near-clean audio, and training operates on full songs (up to $300$~s, i.e.\
${\sim}7.5$k latent frames at $25$~Hz) rather than fixed crops. We initialize from
the Stage~2 checkpoint.

\subsubsection{Stage 4: Vector-Quantized Tokenizer}
\label{sec:tok-stage4-vq}

The final stage discretizes the encoder by inserting a single VQ bottleneck at
an intermediate layer of the $24$-layer Conformer stack. The activation at that
layer is quantized, and the remaining upper layers and the SFT heads then operate
on the quantized stream, forcing the codes to retain the information required for
lyric transcription and spectral reconstruction.

\paragraph{Quantizer}
We insert a single VQ codebook of $32768$ entries under a cosine metric, trained
with a straight-through estimator and a commitment loss.
The quantizer incorporates dedicated mechanisms that keep essentially the entire
codebook active throughout training, so that codebook utilization reaches
$99\%+$ and dead codes are virtually eliminated even at $2^{15}$ entries; the
full code space therefore contributes to the bitrate budget. A single codebook of
size $2^{15}$ at $25$~Hz corresponds to $25$ tokens/s and a bitrate of
$375$~bit/s.

\paragraph{Smooth insertion and staged unfreezing}
Injecting a hard quantizer into a converged encoder is unstable, so we blend the
quantized and continuous paths through a scalar gate,
$h_{\text{out}} = h + \alpha\,(h_q - h)$, where $\alpha$ ramps linearly from $0$
to $1$ over an initial warmup. Concurrently, we first freeze the frontend and all
layers up to and including the insertion point and train only the quantizer, the
upper layers, and the heads, then unfreeze the whole model. The loss is the SFT
objective plus the VQ loss, with the CTC weight lowered to $0.5$ (Mel, chroma, and
VQ weights $1.0$). We initialize from the Stage~3 checkpoint.

\paragraph{Tokenization.}
At inference, extracting codes requires only the frontend, the Conformer layers
up to the insertion point, and the quantizer; the post-VQ layers and all heads are
discarded. Each $40$~ms frame maps to a single integer in $[0,32768)$, giving the
$25$~Hz token stream consumed by downstream models.

\begin{table}[t]
\centering
\small
\caption{Overview of the four-stage training recipe for Qwen-Music-Tokenizer.
All stages share the $0.6$B-parameter Conformer backbone ($24$ layers,
width $1024$, $16$ heads, $25$~Hz latents) and are initialized from the previous
checkpoint.}
\label{tab:tok-recipe}
\begin{tabular}{@{}lllll@{}}
\toprule
Stage & Objective & Attention & Audio length & Init.\ from \\
\midrule
1 & BestRQ masked prediction        & bidirectional & $30$~s crop              & random  \\
2 & BestRQ (causal adaptation)      & causal        & $30$~s crop              & Stage 1 \\
3 & CTC $+$ Mel $+$ chroma SFT      & causal        & full song ($\le\!300$~s) & Stage 2 \\
4 & SFT $+$ VQ                      & causal        & full song ($\le\!300$~s) & Stage 3 \\
\bottomrule
\end{tabular}
\end{table}

\subsection{Qwen-Music-LLM}
\label{sec:qwenmusic_llm}

Qwen-Music-LLM adopts an autoregressive backbone
initialized from a $3$B dense variant of Qwen3.5-Omni~\citep{Qwen35_omni}. It models music in the discrete
semantic-token space produced by Qwen-Music-Tokenizer. Given the output of the
prompt rewriter, including global music tags and structured lyrics, the model
predicts a sequence of Music Semantic Tokens that captures the long-range
composition, vocal phrasing, accompaniment, and section-level musical
development of the target song. These generated tokens are then passed to
Qwen-Music-Render for high-fidelity waveform synthesis.

A central challenge in text-to-music generation is that text prompts typically
specify style and lyrics only at a high level, without explicitly determining the
melodic contour or compositional structure of the song. Directly predicting
full-mixture Music Semantic Tokens from text therefore requires the model to learn composition
and arrangement simultaneously. To address this, we introduce \textbf{Melody-CoT}, a melody-token-based chain-of-thought for explicit melody planning: before generating full-mixture Music Semantic Tokens, the model is
trained to produce an intermediate sequence of melody tokens that describes a
coarse-grained vocal melody contour. This intermediate melody plan encourages
the LLM to first resolve the compositional structure of the song and then render
that plan into the full Music Semantic Token sequence.

Melody-CoT also provides a unified interface for text-to-music and cover song
generation. For text-to-music generation, Qwen-Music-LLM conditions only on tags and
lyrics and directly generates Music Semantic Tokens, optionally producing the melody plan
as an intermediate step. For cover song generation, melody tokens extracted
from a reference song are inserted into the prefix together with the target tags
and lyrics. The model then generates new full-mixture Music Semantic Tokens that follow the
reference melody contour while allowing timbre, arrangement, and genre to be
controlled by the textual conditions.

\subsubsection{Melody Tokenizer}

The Melody Tokenizer converts a reference vocal pitch contour into a compact
sequence of discrete melody tokens. We first extract a $50$~Hz vocal pitch curve
using RMVPE~\citep{rmvpe}. Since the goal is to preserve the compositional melody rather than
fine-grained singing expression, the pitch curve is downsampled by a factor of
{$8$ with median pooling}, yielding a {$6.25$~Hz melody sequence}. This low frame
rate suppresses local ornaments, vibrato, and other performance-level details
while retaining the coarse contour needed for melody conditioning.

To prevent the melody condition from leaking
absolute pitch range, key, or singer-dependent timbre information, we use a
\textbf{relative} MIDI representation. Specifically, we convert the downsampled pitch values from Hertz to MIDI pitch values and then compute the median MIDI value
over all voiced frames in the sequence and subtract it from every voiced frame.
The resulting relative semitone offsets are clipped to a fixed range and mapped
to a {$256$-entry vocabulary}, with one dedicated token reserved for unvoiced
frames. As a result, the melody tokens encode the shape of the vocal melody but
discard most absolute-register information, encouraging the generated song to
obey the target tags and lyrics for style, timbre, and arrangement.

We also explored chromagram-based representations~\citep{musicgen,vevo2} for melody tokenization.
However, our preliminary experiments indicate that chromagrams often retain
substantial information beyond the leading melody, including residual background
and acoustic details of the arrangement. Such extra information can make the conditioning signal
too strong and reduce the model's ability to follow the global music tags. We therefore
adopt pitch-contour-based melody tokens as a cleaner representation of the
reference melody.

Formally, let $\mathbf{f}=(f_{1},\ldots,f_{T})$ denote the
RMVPE pitch curve at $50$~Hz. We first apply $8{\times}$ median-pooling
downsampling,
\begin{equation}
\tilde{\mathbf{f}}=\operatorname{MedianPool}_{8}(\mathbf{f}),
\end{equation}
where unvoiced frames are ignored when computing the median and a pooled frame
is set to $0$ if the corresponding window is mostly unvoiced. The melody token
$z_i$ is then computed as
\begin{equation}
\begin{aligned}
\bar m
&= \operatorname{median}\{
\operatorname{round}(\operatorname{hz2midi}(\tilde f_i))\mid \tilde f_i>0\},\\
z_i
&=
\begin{cases}
\operatorname{clip}(
\operatorname{round}(\operatorname{hz2midi}(\tilde f_i))-\bar m,\,-127,\,127)
+ 127, & \tilde f_i>0,\\
255, & \tilde f_i=0,
\end{cases}
\end{aligned}
\label{eq:melody-tokenizer}
\end{equation}
where $\tilde f_i$ is the downsampled pitch value, $\bar m$ is the global voiced
MIDI median, and $z_i\in[0,255]$ is the final melody token.

\subsubsection{Melody-CoT Training}

Qwen-Music-LLM is trained with a {next-token prediction objective} over sequences
that may contain text tokens, melody tokens, and Music Semantic Tokens. The
text portion, consisting of tags and lyrics, is used as conditioning context.
Loss is applied to {both the Melody-CoT region and the final Music Semantic Token region},
so the model learns not only to generate realistic Music Semantic Tokens, but also to
construct an explicit melody plan from textual conditions.

We mix multiple sequence patterns during training to support different inference
requirements within a single model. The plain text-to-music pattern is
\[
    [\mathrm{text},\ \mathrm{music\ semantic\ tokens}],
\]
where the model directly generates Music Semantic Tokens from tags and lyrics. The
melody-planning patterns insert Melody-CoT before the Music Semantic Token sequence:
\[
    [\mathrm{text},\ \mathrm{section\text{-}level\ melody},\ \mathrm{music\ semantic\ tokens}]
\]
and
\[
    [\mathrm{text},\ \mathrm{unique\text{-}section\text{-}level\ melody},\
    \mathrm{music\ semantic\ tokens}].
\]
Balancing these patterns allows the model to retain strong text-to-music
capability while also learning to use explicit melody conditions for cover song
generation.

The \textbf{section-level} Melody-CoT pattern writes melody tokens for vocal
sections that contain lyrics. Each melody segment is preceded by its section
label, such as \texttt{[verse]} or \texttt{[chorus]}, and non-vocal sections
such as \texttt{intros}, \texttt{instrumental breaks}, \texttt{outros}, and \texttt{silence} are omitted from the
Melody-CoT sequence. This design weakens the melody condition compared with a
full-song pitch trace and avoids exposing the exact full-song duration through
the length of the melody sequence.

The \textbf{unique-section-level} pattern further reduces over-conditioning. Instead of
including every occurrence of a repeated section label, it groups candidate
segments by section type and samples one representative melody segment for each
unique label during training. For example, multiple chorus occurrences are
represented by one sampled chorus melody. This shortens the Melody-CoT sequence,
improves data diversity across epochs, and encourages the model to use melody as
a reusable compositional guide rather than copying the reference song frame by
frame.

\subsection{Qwen-Music-Render}

\begin{figure}[tbh]
    \centering
    \includegraphics[width=0.9\textwidth]{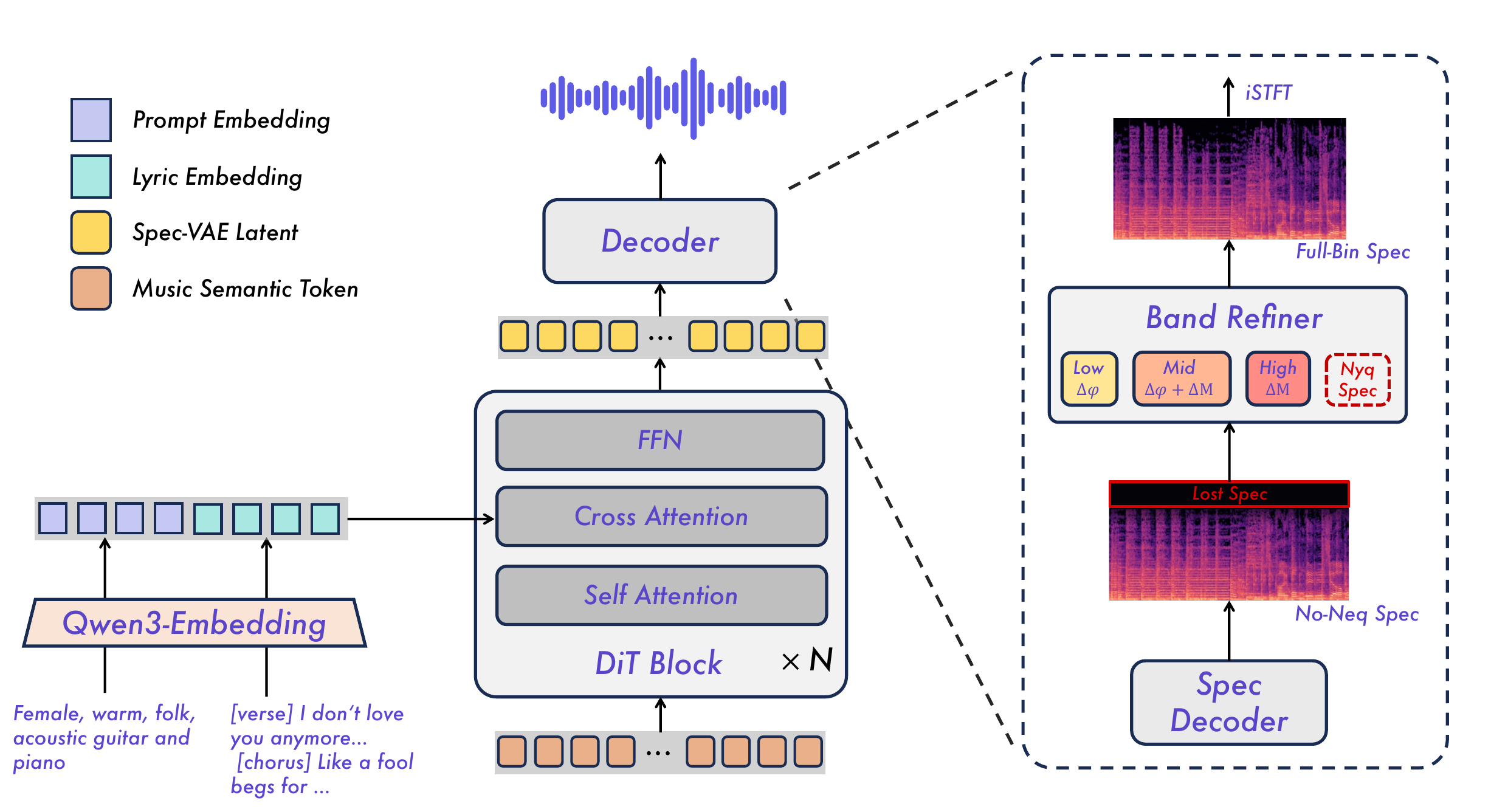}
    \caption{Overview of Qwen-Music-Render. Music Semantic Tokens and text
    conditions guide a semantic-conditioned DiT that predicts
    acoustic latents. Spec-VAE decodes the latents into complex spectrograms,
    and the Band-Mode Refiner corrects band-dependent magnitude and phase details
    before inverse STFT synthesis to 48~kHz stereo waveforms.}
    \label{fig:render_architecture}
\end{figure}

The language model captures long-range musical structure in a compact
discrete token space. We introduce \emph{Qwen-Music-Render}, a three-stage neural
render that turns this discrete output into a high-fidelity waveform.
The first stage is a \emph{Diffusion Transformer} (DiT)~\citep{peebles2023dit,liu2023flow,lipman2023flow} that, conditioned
on the discrete tokens together with the textual musical tags and lyrics, produces a continuous latent through conditional flow
matching. The second stage is \emph{Spec-VAE}, whose Spec Decoder inverts
the latent sequence back to a coarse complex STFT. The
third stage is a \emph{Band-Mode Refiner} that predicts
frequency-adaptive magnitude and phase residuals on top of the decoder
output; a final inverse STFT produces the 48~kHz stereo waveform.

\subsubsection{Diffusion Transformer}
\label{sec:hd_render_dit}

The Music Semantic Tokens generated by the language model provide a discrete representation of the song’s global structure, section ordering, and vocal/instrumental arrangement. Conditioned on these tokens, the Diffusion Transformer (DiT) predicts the latent representation, which is subsequently decoded by the downstream Spec Decoder into a listenable spectrogram.

\paragraph{Architecture}
Our backbone is a $1.3$~B DiT operating on continuous latents. It comprises 32 Transformer blocks with hidden dimension $d = 1024$, 24 attention heads, and a feed-forward expansion ratio of 8. Rotary positional embeddings (RoPE)~\citep{su2024roformer} are applied to the queries and keys in self-attention. Each block consists of a self-attention layer, a cross-attention layer that attends to the auxiliary conditioning context described below, and a feed-forward layer, all modulated by AdaLN parameters derived from the diffusion timestep.
At each forward pass, the network takes as input a noisy latent $\phi_t \in \mathbb{R}^{T \times m}$ with $m = 128$, together with a frame-aligned token sequence $\mathbf{c} \in \{1, \ldots, V\}^T$ generated by the language model at 25~Hz. 

\paragraph{Conditioning}

In Qwen-Music-Render, we further provide the DiT with the rewritten textual condition produced by the prompt rewriter, in addition to the Music Semantic Tokens generated by Qwen-Music-LLM.
We find that this additional textual conditioning improves audio quality and helps the renderer better follow the intended song specification.
Specifically, we use a frozen Qwen3-Embedding-0.6B~\citep{zhang2025qwen3} text encoder to extract features from the textual condition, and inject them into the DiT through cross-attention.
The song-description tokens are projected to the DiT hidden width with a bias-free linear layer, while the lyric tokens are further processed by a 6-layer RoPE Transformer encoder trained jointly with the DiT.
The resulting description and lyric representations are concatenated into a single cross-attention context $\mathcal{C}$ attended by each DiT block.
To enable classifier-free guidance (CFG) at inference time, we introduce a learnable null context $\mathcal{C}_{\emptyset}$ for the text-conditioning branch.
During CFG inference, the conditional pass uses both the Music Semantic Tokens and the full text context $\mathcal{C}$, while the unconditional pass keeps the Music Semantic Tokens unchanged and replaces $\mathcal{C}$ with $\mathcal{C}_{\emptyset}$.

\paragraph{Training details}
We train the model in two stages. First, the DiT is pretrained on a large-scale music corpus. It is then supervised fine-tuned (SFT) on a smaller set of high quality recordings, which further improves output audio quality. Training samples are limited to tracks of up to 6~minutes, corresponding to at most 9,000 frames at 25~Hz. The cross-attention context is truncated to 256 tokens for the song description and 1536 tokens for the lyrics. We use AdamW for optimization, together with a linear warm-up schedule and gradient clipping at $1.0$.

\subsubsection{Spec-VAE \& Refiner}
\label{sec:hd_render_decoder}

\paragraph{Architecture}
Spec-VAE follows the 2D-conv architecture of SpectroStream~\citep{li2025spectrostream}, with discrete residual vector-quantized tokens replaced by a continuous spectral latent to match the latent diffusion framework in Section~\ref{sec:hd_render_dit}.
Stereo audio at 48~kHz is transformed by STFT into a complex spectrogram $\mathbf{X} \in \mathbb{C}^{2 \times 480 \times T}$.
A 7-block Spec Encoder compresses the spectrogram into 128-dimensional latents at 25~Hz, achieving a total compression ratio of $192\times$, while a mirrored Spec Decoder reconstructs a coarse complex spectrogram for iSTFT synthesis.
Stereo channels are handled with delayed fusion in the encoder and early splitting in the decoder.
To further improve reconstruction quality, we attach a lightweight Band-Mode Refiner after the decoder to correct residual spectral artifacts.

\begin{figure}[t]
  \centering
  \includegraphics[width=0.8\columnwidth]{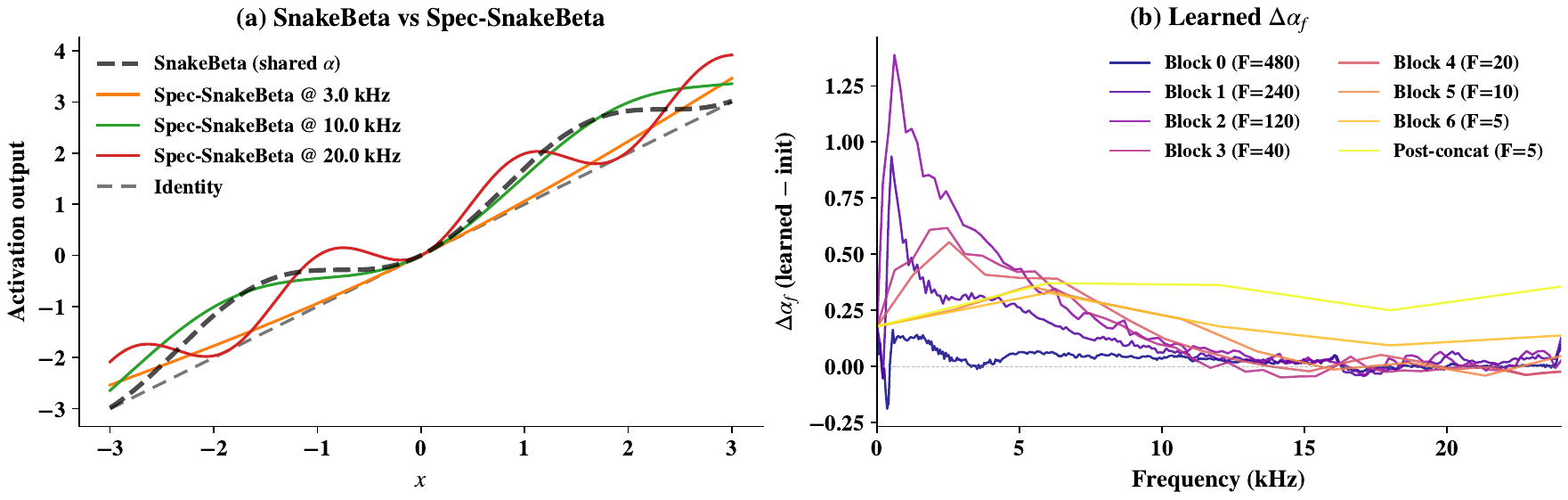}
  \caption{Spec-SnakeBeta analysis.
  \textbf{(a)}~Activation curves: the standard SnakeBeta (black dashed)
  applies a single shared $\alpha$ to all frequency bins; Spec-SnakeBeta
  adapts the periodic modulation per frequency, producing distinct
  nonlinear shapes at 0.5, 3, 10, and 20\,kHz.
  \textbf{(b)}~Deviation of learned $\alpha_f$ from log-frequency
  initialization ($\Delta\alpha_f = \alpha_\text{learned} -
  \alpha_\text{init}$) across encoder layers---positive values indicate
  the network increases periodic modulation beyond the physical prior,
  with the strongest deviations concentrated in 0--5\,kHz.}
  \label{fig:snakebeta2d_analysis}
\end{figure}

\paragraph{Spec-SnakeBeta}
We introduce \emph{Spec-SnakeBeta}, a frequency-aware extension of the SnakeBeta activation used in BigVGAN~\citep{lee2023bigvgan}.
While the original SnakeBeta uses channel-wise learnable parameters, Spec-SnakeBeta parameterizes $\alpha$ and $\beta$ along the frequency axis, exploiting the fact that each STFT bin corresponds to a fixed physical frequency:
\begin{equation}
  \text{Spec\mbox{-}SnakeBeta}(\mathbf{x})_{b,c,t,f}
    = x_{b,c,t,f}
    + \frac{1}{\exp(\beta_f) + \epsilon}\,
      \sin^2\left(x_{b,c,t,f} \cdot \exp(\alpha_f)\right).
  \label{eq:snakebeta_2d}
\end{equation}
We initialize $\alpha_f$ with a log-frequency prior,
\begin{equation}
  \alpha_f = \log\left(\frac{\mathrm{freq}_f}{\bar{f}} + \epsilon\right),
  \qquad
  \bar{f} = \frac{1}{F}\sum_{f} \mathrm{freq}_f,
  \label{eq:sub_alpha_init}
\end{equation}
which encourages stronger periodic modulation at higher frequencies.
As shown in Figure~\ref{fig:snakebeta2d_analysis}, the learned parameters preserve this frequency-dependent structure while adapting most strongly in the low-to-mid frequency range.
This frequency-dependent parameterization provides the activation prior used by the Spec-VAE decoder.

\paragraph{Band-Mode Refiner}
The Band-Mode Refiner is a lightweight ConvNeXt-1D module~\citep{liu2022convnext} that refines the decoder's complex spectrogram output.
Motivated by the complementary error patterns of magnitude and phase across frequency bands, it applies band-specific corrections:
phase-only correction in the low-frequency band, joint magnitude and phase correction in the mid-frequency band, and magnitude-only correction in the high-frequency band.
This design improves spectral fidelity while preserving stable optimization: at initialization, all refiner weights are zero, so the refined output is exactly equal to the decoder output.

\paragraph{Training Objective and Discriminators}
The decoder system is trained with a combination of multi-resolution STFT loss~\citep{yamamoto2020parallel}, IF/GD phase loss~\citep{wang2025earVAE}, LSGAN adversarial loss with feature matching~\citep{mao2017lsgan,larsen2016vaegan}, and KL regularization.
We further adopt K-weighting perceptual filtering~\citep{itu2015bs1770}, MSLR stereo decomposition~\citep{wang2025earVAE}, adaptive log-magnitude normalization from SAME~\citep{liao2025same}, and mixed-scale spectral losses from SpectroStream~\citep{li2025spectrostream}.
For adversarial training, we use complementary STFT, CQT, and spectral-domain discriminators to provide both waveform-level and spectral-domain feedback.

\paragraph{Three-Stage Training Pipeline}
As summarized in Table~\ref{tab:decoder_train_hyperparams}, training proceeds in three stages.
Stage~1 performs reconstruction-only Spec-VAE pretraining.
Stage~2 introduces waveform-domain adversarial training.
Stage~3 freezes the Spec-VAE encoder--decoder and trains the Band-Mode Refiner with both waveform and spectral discriminators.
This staged recipe stabilizes training under the $192\times$ compression setting and progressively improves fine-grained acoustic detail.


\begin{table}[t]
    \centering
    \caption{Spec-VAE and refiner training configuration across three stages.
    Stage~1 is reconstruction-only pretraining.
    Stage~2 introduces waveform-domain adversarial training.
    Stage~3 freezes the encoder--decoder and trains the refiner with
    waveform~(\textbf{W}) and spectral~(\textbf{S}) discriminators.}
    \label{tab:decoder_train_hyperparams}
    \small
    \begin{tabular}{@{}llccc@{}}
        \toprule
        Category & Param & Stage 1 & Stage 2 & Stage 3 \\
        \midrule
        \multirow{4}{*}{Optimizer}
          & Generator optim.          & Muon          & Muon                  & AdamW \\
          & Generator LR              & $10^{-4}$     & $1.5\times10^{-4}$    & $1.5\times10^{-4}$ \\
          & Disc optim                & ---           & Muon                  & AdamW \\
          & Disc LR                   & ---           & $3\times10^{-5}$      & $10^{-6}$ (W) / $10^{-5}$ (S) \\
        \midrule
        \multirow{4}{*}{Schedule}
          & Batch size / GPU          & 4             & 1                     & 1 \\
          & Segment length            & 1.28\,s       & 1.28\,s               & 1.28\,s \\
          & Disc warmup step          & ---           & 0                     & 0 (W) / 15k (S) \\
          & G\!:\!D interleave        & ---           & 1\!:\!1               & 4\!:\!1 \\
        \midrule
        \multirow{3}{*}{Architecture}
        & STFT disc scales & --- & 8 & 8 \\
        & CQT disc & \xmark & \cmark & \cmark \\
        & Spectral disc & \xmark & \xmark & \cmark \\
        \midrule
        \multirow{5}{*}{Loss weights}
          & $\lambda_{\text{STFT}}$                        & 1.0  & 1.0       & 0.5 \\
          & $\lambda_{\text{IF\_GD}}$                      & 0    & 0         & 0.1 \\
          & $\lambda_{\text{KL}}$                          & $10^{-6}$ & $10^{-6}$ & 0 \\
          & $\lambda_{\text{time-GAN}}~(Adv / FM)$       & ---  & 1.0 / 100 & 0.175 / 20 \\
          & $\lambda_{\text{spec-GAN}}~(Adv / FM)$       & ---  & ---       & 0.5 / 35 \\
        \bottomrule
    \end{tabular}
\end{table}

\subsubsection{Acoustic Data Quality Pipeline}
\label{sec:acoustic_pipeline}

We train Qwen-Music-Render only on files whose acoustic evidence matches a
genuine high-fidelity music target. Raw collections often contain lossless
containers transcoded from MP3, duplicated-mono ``stereo'' tracks, clipped or
over-limited masters, and upsampled files whose high-frequency band is empty.
We therefore apply the automated pipeline
summarized in Table~\ref{tab:pipeline_rules} before renderer training and
evaluation.

\begin{table*}[tbh]
  \centering
  \scriptsize
  \setlength{\tabcolsep}{3.5pt}
  \renewcommand{\arraystretch}{1.10}
  \caption{Metric-level checklist used by the acoustic pipeline.
  Each row pairs a rule-engine identifier with a brief description
  and its pass/flag condition.}
  \label{tab:pipeline_rules}
  \begin{tabular}{@{}p{1.8cm}p{3.0cm}p{5.5cm}p{4.3cm}@{}}
    \toprule
    \textbf{Check} & \textbf{Metric ID} & \textbf{Evidence} & \textbf{Condition} \\
    \midrule
    \multirow{2}{1.8cm}{Meta \& encoder}
      & \texttt{encoder\_tag}     & Encoder string in container/stream tags        & flag if \texttt{lavf/lavc/ffmpeg/lame} \\
      & \texttt{mutagen\_tags}    & Metadata sweep for conversion traces           & flag if \texttt{transcode/convert} \\
    \midrule
    \multirow{5}{1.8cm}{Bitrate}
      & \texttt{has\_non\_monotonic\_pts} & PTS monotonicity              & pass if \texttt{false} \\
      & \texttt{bitrate\_utilization}     & Realized / nominal bitrate    & pass if $\geq0.6$ \\
      & \texttt{top10\_bitrate\_mean}     & Complex-segment mean bitrate  & pass if $\geq1.2\times$\,mean \\
      & \texttt{vbr\_variation}           & Bitrate coefficient of variation & pass if $\sigma/\mu\geq0.15$ \\
      & \texttt{quality\_tier}            & Bitrate-based quality level & HiRes/High/Med/Low/Trash \\
    \midrule
    \multirow{4}{1.8cm}{Loudness}
      & \texttt{lufs\_i}            & Integrated loudness (EBU R128)          & $[-25,-5]$\,LUFS \\
      & \texttt{true\_peak}         & dBTP level                              & $[-1,+2]$\,dBFS \\
      & \texttt{lra}                & Loudness range                          & $[3,15]$\,LU \\
      & \texttt{plr}                & Peak-to-loudness ratio                  & $\geq6$\,LU \\
    \midrule
    \multirow{1}{1.8cm}{Fake stereo}
      & \texttt{lr\_diff\_zero\_ratio} & Fraction of zero $|L-R|$ samples & pass if $<0.99$ \\
    \midrule
    \multirow{3}{1.8cm}{Freq cutoff}
      & \texttt{cutoff\_ratio}     & Cutoff / Nyquist freq ratio             & ranking from $0.5 \to 1.0$ \\
      & \texttt{drop\_db}          & Energy gap around $f_c$                 & hard $\geq40$\,dB; suspicious $\geq20$\,dB \\
      & \texttt{valid\_segments}   & Non-silent interior FFT windows         & pass if $>0$ \\
    \bottomrule
  \end{tabular}
\end{table*}

\paragraph{Noise-floor-aware cutoff detection}
Container metadata alone cannot distinguish true high-resolution content from
an upsampled or lossy-transcoded file. The key idea is to estimate the noise
floor from track \emph{edges} and detect the spectral cutoff from
\emph{interior} segments, keeping the two estimates independent.

We reserve the first and last 10\% of the mono waveform as edge segments and
compute Hann-windowed $8192$-point RFFTs over non-overlapping windows. After
peak-normalizing and averaging their magnitude spectra, the empirical noise
floor is
\begin{equation}\label{eq:noise_floor}
  \hat\eta =
  20\log_{10}\bigl(\mathrm{median}(\bar{\mathbf{M}}_\mathrm{edge}[\text{top 10\% bins}])\bigr).
\end{equation}

For the cutoff, we process five equally spaced interior windows that skip
near-silent regions and average the resulting spectra. The detected cutoff
$f_c$ is the highest FFT bin whose mean magnitude remains above $-80$~dB
relative to the peak. We then measure a local energy drop across a $\pm2$~kHz
band around $f_c$:
\begin{equation}\label{eq:drop_db}
  \Delta_\mathrm{dB} =
    \max\bar{\mathbf{M}}_\mathrm{int}[f_c{-}2\,\text{kHz}:f_c]
    -
    \max\bar{\mathbf{M}}_\mathrm{int}[f_c:f_c{+}2\,\text{kHz}],
\end{equation}
where the computation replaces the post-cutoff maximum with $\hat\eta$ when no
reliable content exists above $f_c$. The final label depends on both the cutoff
ratio $f_c/f_\mathrm{Nyq}$ and the drop magnitude. The rule marks a file as
\texttt{hard\_cutoff} when the ratio falls below $0.85$ and the drop reaches
$\Delta_\mathrm{dB}\geq 40$~dB, or $\geq 25$~dB near $\hat\eta$. It marks a
file as \texttt{suspicious} when the ratio is below $0.90$ with
$\Delta_\mathrm{dB}\geq 20$~dB. Files that meet neither condition are labeled
\texttt{natural}.

\begin{figure*}[t]
  \centering
  \includegraphics[width=\textwidth]{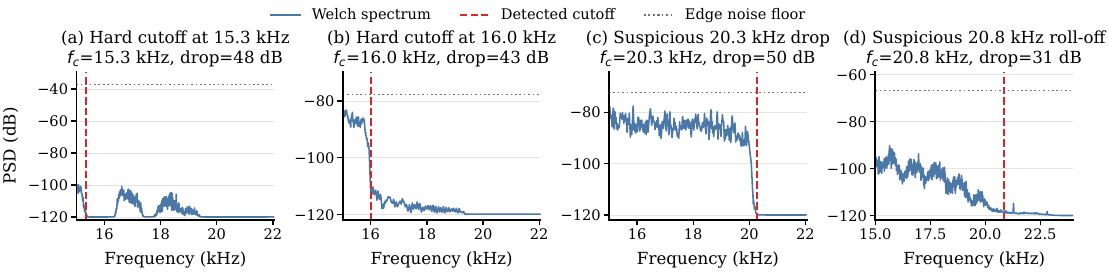}
  \caption{High-frequency cutoff detector examples. The blue
  curve is the Welch spectrum of a 10~s verification segment, the red dashed
  line is the detected cutoff $f_c$, and the gray dotted line is the independently
  estimated edge noise floor.}
  \label{fig:acoustic_cutoff_examples}
\end{figure*}

Figure~\ref{fig:acoustic_cutoff_examples} visualizes four representative
outcomes. The first two panels show classic lossy-to-lossless signatures: a
sharp wall well below Nyquist and post-cutoff energy collapsed to the noise
floor. The last two panels are closer to Nyquist and therefore require the
noise-floor-aware drop test; the detector flags them only when the roll-off is
both low enough in frequency and steep enough in level.

\section{Training}
\label{sec:train}



The remaining training pipeline targets Qwen-Music-LLM. Renderer-specific
acoustic filtering appears in Section~\ref{sec:acoustic_pipeline}; here we train
the backbone LLM through quality-aware pre-training and progressive
post-training. 

\subsection{Quality-Graded Pre-training Curriculum}
\label{sec:pretraining}

%
%

To effectively exploit large-scale music corpora with heterogeneous quality distributions, Qwen-Music adopts a quality-graded pre-training curriculum for the backbone LLM.
Instead of simply discarding all lower-quality data, we organize the corpus into quality levels and schedule them progressively, allowing the model to first learn broad musical coverage and then concentrate on higher-quality generation.

Following the multi-aspect song-quality assessment protocols of SongBench~\citep{wu2026songbench} and SongEval~\citep{yao2025songeval}, we train an internal MOS-based reward model.
The model predicts scores that are used to assess and rank the overall musical quality of each audio sample.
Within each genre, we rank samples and partition them into seven quality buckets (Q1--Q7)
using percentile thresholds at 90\%, 75\%, 50\%, 25\%, 5\%, and 1\%. Q1 denotes the
highest-quality bucket, and Q7 denotes the
lowest-quality bucket. This genre-normalized stratification reduces bias toward
dominant styles and balances quality modeling across musical categories. We
remove Q7 and use the remaining Q1--Q6 subsets for model training.

As described in Section~\ref{sec:qwenmusic_llm}, Qwen-Music-LLM is initialized from a $3$B dense variant of Qwen3.5-Omni~\citep{Qwen35_omni}, whose language-model backbone provides useful prior knowledge of language and music.
We then train the model with a three-stage curriculum that gradually moves from broad data coverage to high-quality refinement.


\paragraph{Stage 1: General pre-training}
In the first stage, we train on Q3--Q6 data, which constitute the majority of the training corpus.
The objective is to learn a general mapping from text conditions, including lyrics and musical tags, to music semantic tokens.
To improve both robustness and generation quality, we use a dynamic quality sampling strategy in which lower-quality data is emphasized early in training and gradually replaced by higher-quality data as training progresses.
We further apply language balancing to mitigate cross-lingual data imbalance and include 20\% instrumental music to strengthen accompaniment modeling.


\paragraph{Stage 2: Annealing}
In the second stage, we train on Q2-level data with a learning-rate decay schedule.
This stage serves as a quality consolidation phase, helping the model transition from broad data coverage to more structured and stable music generation.
It improves musical structure, coherence, and overall audio quality while reducing artifacts introduced by noisier data in the first stage.


\paragraph{Stage 3: High-quality refinement}
In the final stage, we focus on Q1 data and additionally incorporate carefully selected high-quality samples from the full corpus.
This stage further improves musicality, controllability, and instruction following.
We maintain balanced genre and language distributions to avoid overfitting to dominant styles and to ensure stable generation under diverse control conditions.

\subsection{Multi-Stage Post-training Alignment}
\label{sec:posttraining}


After pre-training, we further align Qwen-Music-LLM with human musical preferences and user instructions through a multi-stage post-training pipeline.
The pipeline optimizes two complementary objectives: improving musicality, measured by the internal musicality MOS predictor described above, and enhancing instruction following, including genre control and lyric following, with rewards computed by Qwen3.5-Omni~\citep{Qwen35_omni} and Qwen3-ASR~\citep{shi2026qwen3}.
We organize post-training from stable to exploratory: supervised learning first establishes a clean generation prior, offline preference optimization then improves alignment in a controlled manner, and online policy optimization finally unlocks further gains in musicality and audio quality.


\paragraph{Phase I: Supervised Cold Start}
We first perform supervised fine-tuning on carefully curated high-quality data.
This stage selects samples that are both aligned with human preference and reliably annotated with control information, especially genre tags and lyrics.
By emphasizing high-quality music examples with accurate style labels and lyric annotations, the model establishes a clean generation prior with strong controllability before reward-based optimization begins.

\paragraph{Phase II: Iterative Offline Preference Alignment}
Building on the SFT model, we perform iterative offline alignment with Direct Preference Optimization (DPO)~\citep{rafailov2024direct}.
In each iteration, the current policy generates rollouts under diverse prompts and control conditions, and the musicality MOS predictor together with instruction-following reward models score these outputs.
We then construct preference pairs from the scored candidates and optimize Qwen-Music-LLM with DPO.
Repeating this rollout--rewarding--pair construction--optimization loop progressively improves musicality and controllability while producing a policy that is more stable and better aligned for subsequent online optimization.


\paragraph{Phase III: Online Musicality Optimization}
Finally, we apply on-policy Group Sequence Policy Optimization (GSPO)~\citep{zheng2025group}.
Compared with offline DPO, GSPO enables stronger exploration by updating the policy directly from on-policy samples and sequence-level rewards.
Because the model has already acquired reliable generation ability and instruction-following behavior in the first two phases, this stage can focus on fine-grained musicality and audio-quality improvement rather than correcting fundamental alignment issues.


Together, these three phases provide a stable optimization path for Qwen-Music-LLM, yielding stronger musicality, better controllability, and more reliable instruction following.

\section{Evaluation}
\label{sec:experiment}



Qwen-Music is evaluated on two core tasks: text-to-music generation and cover song generation.
In blind A/B preference tests conducted by professional human raters (Figure~\ref{fig:ab_preference}), Qwen-Music outperforms MiniMax Music 2.6, MiniMax Music 2.5+, Mureka V8, and Suno V5, while remaining competitive with Suno V5.5.
Genre-wise Bradley--Terry analysis further shows strong preference across diverse musical styles (Table~\ref{tab:genre_bt}), and the Artificial Analysis leaderboard provides an external reference point for its performance among leading English vocal music generation systems (Figure~\ref{fig:jazzcat_aa}).
Beyond full-song generation, we also evaluate Qwen-Music-Render through generative rendering ablations, acoustic reconstruction benchmarks, and refiner-specific analyses.
Together, these evaluations show that Qwen-Music combines strong musicality, controllability, lyric intelligibility, cover-song melody preservation, and high-fidelity stereo rendering.

\subsection{Evaluation of Music Generation}

We evaluate Qwen-Music on two music generation tasks: \textbf{text-to-music generation} and \textbf{cover song generation}.
Text-to-music generation measures the ability to create complete songs from a user's natural-language prompt or request.
Cover song generation measures the ability to reinterpret a reference song or melody according to the user's desired target style and vocal direction.

\paragraph{Baseline Systems}
For text-to-music generation, we compare Qwen-Music against five state-of-the-art proprietary music generation systems: Suno V5.5, Suno V5, Mureka V8, MiniMax Music 2.6, and MiniMax Music 2.5+.
For cover song generation, we retain the leading proprietary services that support this capability: Suno V5.5, Suno V5, and MiniMax Cover.
Since the Suno systems do not accept real-world songs as reference inputs, comparisons with Suno are limited to the AI-generated reference set.

\paragraph{Evaluation Metrics}
We use complementary objective metrics to evaluate musicality, audio quality, controllability, intelligibility, and melody preservation.
For general musical quality, SongBench~\cite{wu2026songbench} scores melody, arrangement, musicality, vocal quality, instrumental quality, mixing, and structure, with higher values indicating better performance.
For text-to-music generation, we also report SongEval~\cite{yao2025songeval} and AudioBox-Aesthetic~\cite{tjandra2025meta} as additional music-aesthetic evaluation frameworks.
For lyric intelligibility and lyric following, Qwen3-ASR~\citep{shi2026qwen3} transcribes the generated audio and we compute the Phoneme Error Rate (PER) against the input lyrics; lower PER is better, and we report PER after multiplying by 100.
For tag following, Gemini 3.1 Pro~\cite{comanici2025gemini} rates each generated song on a 1--10 scale across genre, mood, instrument, vocal gender, and vocal timbre, where higher scores indicate better adherence to the target musical tags.
For cover song generation, we further measure reference-melody preservation using Melody MAE.
Specifically, we convert both the generated cover and the reference song into relative MIDI representations following Eq.~\ref{eq:melody-tokenizer} in Section~\ref{sec:architecture}, align the two melody sequences with dynamic time warping, and compute the mean absolute error in semitones.
Lower Melody MAE indicates stronger preservation of the reference melody.

\subsubsection{Text-to-Music Generation}
\begin{table*}[t]
    \centering
    \caption{
      Genre-wise Bradley--Terry ratings from blind A/B preference tests conducted by professional human raters.
      Ratings are computed independently within each genre and should be interpreted only for within-genre comparison.
      Higher ratings indicate stronger preference.
      Best results are highlighted in \textbf{bold}, and second-best results are \underline{underlined}.
    }
    \label{tab:genre_bt}
    \resizebox{0.9\textwidth}{!}{
    \begin{tabular}{lcccccc}
    \toprule
    \textbf{Genre} 
    & \textbf{Qwen-Music} 
    & \textbf{Suno V5.5} 
    & \textbf{Suno V5} 
    & \textbf{Mureka V8} 
    & \makecell{\textbf{Minimax}\\\textbf{Music 2.6}} 
    & \makecell{\textbf{Minimax}\\\textbf{Music 2.5+}} \\
    \midrule
    Electronic / EDM 
        & \textbf{1140} & 880 & 853 & \underline{938} & 759 & 866 \\
    Hip-Hop / Rap 
        & \underline{1020} & \textbf{1205} & 1018 & 898 & 866 & 930 \\
    Jazz \& Blues 
        & \textbf{1112} & 961 & \underline{968} & 202 & 868 & 930 \\
    Metal \& Hard Rock 
        & 1039 & 880 & \underline{1058} & \textbf{1070} & 859 & 942 \\
    Pop 
        & \underline{1020} & 1000 & 980 & \textbf{1050} & 977 & 880 \\
    Punk \& Hardcore 
        & \textbf{1147} & 863 & 759 & 918 & \underline{1120} & 202 \\
    R\&B / Soul 
        & \textbf{1056} & \underline{1032} & 950 & 1000 & 853 & 880 \\
    Rock 
        & \textbf{1116} & 859 & 782 & 961 & 720 & \underline{1027} \\
    \bottomrule
    \end{tabular}
    }
\end{table*}

\paragraph{Subjective Evaluation}
To complement the automatic evaluation, we conduct a blind A/B preference test between Qwen-Music and competing systems, with judgments collected from professional human raters.
We use the same generated audio samples as in the objective evaluation.
For each prompt and each comparison system, the audio generated by Qwen-Music and the corresponding baseline are paired and presented in random order.

A total of 50 professional human raters with music creation or production experience are recruited.
In each trial, raters listen to two anonymized audio samples without knowing their model identities or generation sources, and select the one they prefer based on overall listening experience.
To reduce positional bias, the order of the two samples is randomized for each trial.
Each A/B pair is independently judged by three different experts, and final preference rates are aggregated over all expert votes.


As shown in Figure~\ref{fig:ab_preference}, Qwen-Music is preferred over all comparison systems on average.
It achieves a 59.1\% win rate against MiniMax Music 2.5+, 66.7\% against MiniMax Music 2.6, 58.3\% against Mureka V8, and 55.4\% against Suno V5.
The closest comparison is against Suno V5.5, where Qwen-Music still obtains a slight preference advantage of 50.3\% versus 49.7\%, indicating comparable subjective quality to this strong commercial baseline.



To further analyze robustness across musical styles, we group the test samples according to their genre tags and compute genre-wise Bradley--Terry ratings from the A/B preference outcomes.
Ratings are estimated independently within each genre based on pairwise outcomes between Qwen-Music and each comparison system, and therefore should not be compared across different genres.

As shown in Table~\ref{tab:genre_bt}, Qwen-Music achieves the highest Bradley--Terry rating in five out of eight genres, including Electronic/EDM, Jazz \& Blues, Punk \& Hardcore, R\&B/Soul, and Rock.
It also ranks second in Hip-Hop/Rap and Pop, and remains competitive in Metal \& Hard Rock.
These results suggest that Qwen-Music maintains strong subjective preference across diverse musical styles, rather than showing advantages only in a narrow set of genres.

Figure~\ref{fig:jazzcat_aa} provides an additional external leaderboard reference.
On the Artificial Analysis Music with Vocals Leaderboard, Qwen-Music participates as \texttt{JazzCat} and ranks in the top tier, reaching third place among leading English vocal music generation systems as shown in the figure.
Together with the professional-rater A/B preference test and genre-wise Bradley--Terry analysis, this result further supports the strong and stable perceptual quality of Qwen-Music.

\paragraph{Objective Evaluation}
As shown in Table~\ref{tab:music_benchmarks}, we conduct a comprehensive objective evaluation on a test set containing 300 Chinese and 300 English samples.
To ensure a fair comparison, all systems are evaluated under identical input conditions, using AI-generated lyrics and musical tags with a balanced distribution across genres.

\begin{table*}[t]
    \centering
    \caption[Objective text-to-music generation results.] {Objective text-to-music generation results on 600 evaluation inputs, each comprising AI-generated lyrics and musical tags.
    Musicality and audio quality are evaluated using SongBench, SongEval, and AudioBox-Aesthetic.
    Tag following is evaluated by Gemini 3.1 Pro.
    Higher is better except for PER.
    Best results are shown in \textbf{bold}, and second-best results are \underline{underlined}.}
    \label{tab:music_benchmarks}
    \resizebox{0.9\textwidth}{!}{
    \begin{tabular}{lcccccc}
    \toprule
    \textbf{Metrics} 
    & \textbf{Qwen-Music} 
    & \textbf{Suno V5.5} 
    & \textbf{Suno V5} 
    & \textbf{Mureka V8} 
    & \makecell{\textbf{Minimax}\\\textbf{Music 2.6}} 
    & \makecell{\textbf{Minimax}\\\textbf{Music 2.5+}} \\
    \midrule
    \multicolumn{7}{c}{\textbf{\textit{SongBench}}} \\
    \midrule
    Melody       & \textbf{7.03} & 6.70 & \underline{7.01} & 6.98 & 6.64 & 6.41 \\
    Arrangement  & \textbf{7.35} & 7.03 & \underline{7.20} & 7.14 & 6.72 & 6.45 \\
    Musicality   & \textbf{6.22} & 5.83 & \underline{6.12} & 6.05 & 5.67 & 5.50 \\
    Vocal        & \textbf{7.44} & 6.95 & 7.37 & \underline{7.38} & 6.99 & 6.89 \\
    Instrumental & \textbf{7.24} & 6.85 & \underline{7.03} & 7.01 & 6.64 & 6.52 \\
    Mixing       & \textbf{7.13} & 6.88 & \underline{7.07} & 6.97 & 6.59 & 6.38 \\
    Structure    & 6.94 & 6.83 & \underline{6.98} & \textbf{7.02} & 6.55 & 6.27 \\
    \midrule
    \multicolumn{7}{c}{\textbf{\textit{SongEval}}} \\
    \midrule
    Coherence    & \textbf{4.55} & 4.28 & 4.44 & \underline{4.49} & 4.33 & 4.28 \\
    Musicality   & \textbf{4.42} & 4.12 & 4.30 & \underline{4.36} & 4.19 & 4.15 \\
    Memorability & \textbf{4.51} & 4.27 & \underline{4.43} & \textbf{4.51} & 4.29 & 4.23 \\
    Clarity      & \textbf{4.45} & 4.17 & 4.33 & \underline{4.37} & 4.23 & 4.17 \\
    Naturalness  & \textbf{4.37} & 4.10 & \underline{4.27} & \textbf{4.37} & 4.16 & 4.08 \\
    \midrule
    \multicolumn{7}{c}{\textbf{\textit{AudioBox-Aesthetic}}} \\
    \midrule
    Content Enjoyment      & \textbf{7.47} & 7.35 & 7.40 & 7.31 & 7.39 & \underline{7.42} \\
    Content Usefulness     & \textbf{7.86} & \underline{7.85} & 7.75 & 7.67 & 7.82 & 7.83 \\
    Production Complexity  & 6.64 & \underline{6.67} & \textbf{6.71} & \underline{6.67} & 6.64 & 6.49 \\
    Production Quality     & \underline{8.15} & 8.08 & 8.08 & 7.95 & \underline{8.15} & \textbf{8.20} \\
    \midrule
    \multicolumn{7}{c}{\textbf{\textit{Tags Following}}} \\
    \midrule
    Genre         & 8.08 & \underline{8.33} & 8.28 & \textbf{8.46} & 8.01 & 7.60 \\
    Moods         & \underline{8.94} & 8.93 & \textbf{8.98} & 8.67 & 8.57 & 8.20 \\
    Instruments   & 8.65 & \underline{8.80} & \textbf{8.89} & 8.66 & 8.15 & 7.79 \\
    Vocal Gender & \textbf{7.85} & 7.54 & 7.58 & \underline{7.77} & 7.57 & 7.48 \\
    Vocal Timbre & \underline{8.68} & \textbf{8.71} & 8.67 & 8.38 & 8.31 & 8.58 \\
    \midrule
    \multicolumn{7}{c}{\textbf{\textit{Intelligibility}}} \\
    \midrule
    PER           & \underline{6.10} & \textbf{4.19} & 6.80 & 9.04 & 6.40 & 6.29 \\
    \bottomrule
    \end{tabular}
    }
\end{table*}

Qwen-Music achieves strong performance across musicality and audio-quality metrics.
Across SongBench, SongEval, and AudioBox-Aesthetic, it obtains the best result in 13 out of 16 evaluated dimensions.
On SongBench, Qwen-Music ranks first in six out of seven dimensions, including melody, arrangement, musicality, vocal quality, instrumental quality, and mixing.
On SongEval, it consistently ranks first across all five dimensions: coherence, musicality, memorability, clarity, and naturalness.
On AudioBox-Aesthetic, Qwen-Music achieves the best results in content enjoyment and content usefulness while remaining competitive on production-related metrics.
These results indicate strong overall generation quality in both composition and rendering.

Qwen-Music also demonstrates competitive controllability and lyric intelligibility.
On tag following, it achieves an average score of 8.44 across five control dimensions, within 0.04 points of the best-performing system on average.
It performs particularly well on fine-grained vocal control, achieving the best result on vocal gender following and ranking second on both mood and vocal timbre control.
Although Mureka V8 and Suno V5 obtain slightly higher scores on genre and instrument control respectively, Qwen-Music remains competitive across all dimensions, indicating balanced controllability across different musical attributes.

For intelligibility, Qwen-Music achieves a PER of 6.10, the second-lowest among all systems.
Overall, these results show that Qwen-Music offers a strong balance among musicality, audio quality, controllability, and lyric intelligibility.

\subsubsection{Cover Song Generation}

\begin{table*}[t]
  \centering
  \scriptsize
  \renewcommand{\arraystretch}{0.84}
  \setlength{\tabcolsep}{4pt}
  \caption{Objective cover song generation results on the AI-generated reference-melody set, containing 100 English and 100 Chinese examples.
  The Qwen-Music (section) column uses section-level melody conditioning, while the Qwen-Music (unique section) column uses unique-section-level melody conditioning.
  Higher is better except for PER and Melody MAE.
  Best results are shown in \textbf{bold}, and second-best results are \underline{underlined}.}
  \label{tab:cover-ai-eval}
  \resizebox{0.8\textwidth}{!}{
  \begin{tabular}{@{}lccccc@{}}
  \toprule
  \textbf{Metric} & \makecell{\textbf{Qwen-Music}\\\textbf{(section)}} & \makecell{\textbf{Qwen-Music}\\\textbf{(unique section)}} & \textbf{Suno V5.5} & \textbf{Suno V5} & \textbf{MiniMax Cover} \\
  \midrule
  \multicolumn{6}{c}{\textbf{\textit{SongBench}}} \\
  \midrule
  Melody & 6.86 & 6.85 & \underline{7.13} & \textbf{7.28} & 6.64 \\
  Arrangement & 7.27 & 7.25 & \underline{7.56} & \textbf{7.61} & 6.70 \\
  Musicality & 6.05 & 6.04 & \underline{6.36} & \textbf{6.37} & 5.72 \\
  Vocal & 7.30 & 7.29 & \underline{7.35} & \textbf{7.47} & 7.14 \\
  Instrumental & \underline{7.31} & 7.26 & 7.28 & \textbf{7.34} & 6.74 \\
  Mixing & 6.96 & 6.94 & \underline{7.29} & \textbf{7.33} & 6.66 \\
  Structure & 6.83 & 6.80 & \underline{7.19} & \textbf{7.27} & 6.56 \\
  \midrule
  \multicolumn{6}{c}{\textbf{\textit{Intelligibility}}} \\
  \midrule
  PER ($\downarrow$) & 19.28 & 17.47 & \textbf{5.77} & \underline{8.46} & 25.35 \\
  \midrule
  \multicolumn{6}{c}{\textbf{\textit{Melody Following}}} \\
  \midrule
  Melody MAE ($\downarrow$) & \textbf{1.48} & \underline{1.80} & 2.00 & 1.87 & 1.89 \\
  \midrule
  \multicolumn{6}{c}{\textbf{\textit{Tag Following}}} \\
  \midrule
  Genre & 5.92 & \underline{7.34} & \textbf{8.27} & 7.14 & 5.42 \\
  Mood & 7.32 & \underline{8.51} & \textbf{8.77} & 7.92 & 7.19 \\
  Instruments & 7.55 & \underline{8.42} & \textbf{9.14} & 8.35 & 7.22 \\
  Vocal Gender & 7.96 & 8.89 & \textbf{9.38} & \underline{9.16} & 6.35 \\
  Vocal Timbre & 7.77 & \underline{8.69} & \textbf{9.23} & 8.53 & 7.88 \\
  \bottomrule
  \end{tabular}}
  \end{table*}
  
  \begin{table*}[t]
  \centering
  \scriptsize
  \renewcommand{\arraystretch}{0.84}
  \setlength{\tabcolsep}{4pt}
  \caption{Objective cover song generation results on the openly licensed music reference set, containing 100 English and 100 Chinese examples.
  The Qwen-Music (section) column uses section-level melody conditioning, while the Qwen-Music (unique section) column uses unique-section-level melody conditioning.
  Higher is better except for PER and Melody MAE.
  Best results are shown in \textbf{bold}, and second-best results are \underline{underlined}.}
  \label{tab:cover-hotsong-eval}
  \resizebox{0.6\textwidth}{!}{
  \begin{tabular}{@{}lccc@{}}
  \toprule
  \textbf{Metric} & \makecell{\textbf{Qwen-Music}\\\textbf{(section)}} & \makecell{\textbf{Qwen-Music}\\\textbf{(unique section)}} & \textbf{MiniMax Cover} \\
  \midrule
  \multicolumn{4}{c}{\textbf{\textit{SongBench}}} \\
  \midrule
  Melody & \underline{6.55} & \textbf{6.57} & 6.18 \\
  Arrangement & \underline{6.84} & \textbf{6.86} & 6.16 \\
  Musicality & \underline{5.75} & \textbf{5.79} & 5.30 \\
  Vocal & \underline{7.07} & \textbf{7.09} & 6.73 \\
  Instrumental & \underline{6.96} & \textbf{7.00} & 6.41 \\
  Mixing & \underline{6.66} & \textbf{6.67} & 6.25 \\
  Structure & \textbf{6.48} & \underline{6.46} & 6.13 \\
  \midrule
  \multicolumn{4}{c}{\textbf{\textit{Intelligibility}}} \\
  \midrule
  PER ($\downarrow$) & \underline{20.20} & \textbf{18.49} & 22.48 \\
  \midrule
  \multicolumn{4}{c}{\textbf{\textit{Melody Following}}} \\
  \midrule
  Melody MAE ($\downarrow$) & \textbf{1.44} & 1.80 & \underline{1.76} \\
  \midrule
  \multicolumn{4}{c}{\textbf{\textit{Tag Following}}} \\
  \midrule
  Genre & \underline{6.90} & \textbf{7.35} & 5.84 \\
  Mood & \underline{7.48} & \textbf{8.05} & 7.05 \\
  Instruments & \underline{8.05} & \textbf{8.30} & 7.23 \\
  Vocal Gender & \underline{8.88} & \textbf{9.18} & 6.80 \\
  Vocal Timbre & \underline{8.21} & \textbf{8.54} & 7.75 \\
  \bottomrule
  \end{tabular}}
  \end{table*}

We evaluate cover song generation under reference-melody conditioning.
Given a reference melody, target musical tags, and lyrics, the model is expected to generate a new song that preserves the reference melody while following the requested style and vocal attributes.
We construct two evaluation sets.
The first uses reference songs generated by AI music models, including Suno V5.5 and Mureka V8, with 100 English and 100 Chinese examples.
The second evaluation set uses openly licensed music as reference material, also with 100 English and 100 Chinese examples.
In the tables, \emph{section-level melody} denotes conditioning with melody tokens for each lyric-bearing section, while \emph{unique-section-level melody} denotes conditioning with one representative melody for each repeated section label.

On the AI-generated reference set (Table~\ref{tab:cover-ai-eval}), section-level Melody-CoT achieves the lowest Melody MAE, indicating stronger direct reference-melody cloning, while unique-section-level conditioning substantially improves tag following over the section-level mode.
Commercial systems remain strong on several general musicality and tag-following metrics.
On the real-world popular-song set (Table~\ref{tab:cover-hotsong-eval}), Qwen-Music outperforms MiniMax Cover across most objective dimensions.
The unique-section-level mode yields better musicality, intelligibility, and tag following on most metrics, whereas the section-level mode gives the best direct melody following.
These results show that the two Melody-CoT modes provide complementary control trade-offs for cover song generation.

\subsection{Evaluation of Qwen-Music-Render}

Qwen-Music-Render performs generative rendering from Music Semantic Tokens and the rewritten textual condition into high-fidelity 48\,kHz stereo waveforms.
This section evaluates two key aspects of this rendering module.
We first study the DiT rendering pipeline, focusing on the effects of rewritten textual conditioning, CFG strategy, and latent backbone choice.
We then evaluate whether the Band-Mode Refiner improves acoustic reconstruction quality when paired with Spec-VAE.
Overall, the results show that rewritten textual conditioning with \textit{text-drop} CFG gives the strongest generation quality, and that the refiner consistently improves spectral, mel-scale, and stereo reconstruction metrics.

\subsubsection{Effect of Text Conditioning and CFG in DiT Rendering}
\label{sec:renderer_eval}

Table~\ref{tab:dit_codec_eval} studies three design choices in the DiT rendering pipeline: the latent backbone used as the prediction target, the rewritten textual condition provided to the renderer, and the null-branch design used for classifier-free guidance (CFG).
``None'' indicates no text conditioning, while frame-aligned LM token conditioning is still provided.

The results show that rewritten textual conditioning provides useful guidance beyond the song-level scaffold carried by LM tokens: under the matched Spec-VAE + LM-token drop setting, musical tags improve over no-text conditioning, and adding lyrics further improves most metrics.
Among CFG variants, \textit{Text-drop} CFG achieves the best overall performance, supporting our design choice in Section~\ref{sec:hd_render_dit}: LM tokens should be retained as the primary rendering scaffold, while musical tags and lyrics serve as complementary semantic guidance.
Under matched text-conditioning and CFG settings, Spec-VAE also generally outperforms Levo~2 VAE~\citep{lei2026levo2stablemelodious}, indicating that a stronger latent representation leads to better final rendering quality.
Overall, the best configuration combines \textbf{Spec-VAE, musical tags + lyrics, and \textit{Text-drop} CFG}.

\begin{table*}[!tbh]
  \centering
  \small
  \caption{DiT rendering ablation on our bilingual EN/ZH set (100 samples each).
  We compare latent backbones, rewritten textual conditioning, and CFG null-branch designs.
  For fair comparison, Spec-VAE results here bypass the Band-Mode Refiner.
  ``None'' indicates no text conditioning, while frame-aligned LM token conditioning is always provided.
  \textit{Text-drop}, \textit{Full-drop}, and \textit{LM-token-drop} respectively remove text conditioning only, both text and LM token conditioning, and LM token conditioning only from the null branch.
  Best results are highlighted in \textbf{bold}, and second-best results are \underline{underlined}.}
  \label{tab:dit_codec_eval}
  \resizebox{\textwidth}{!}{%
  \begin{tabular}{@{}lcccccc@{}}
    \toprule
    \textbf{Latent Backbone}
      & Levo~2 VAE
      & Levo~2 VAE
      & Spec-VAE
      & Spec-VAE
      & Spec-VAE
      & Spec-VAE \\
    \textbf{Text Conditioning}
      & None
      & Tags + Lyrics
      & Tags only
      & Tags + Lyrics
      & Tags + Lyrics
      & Tags + Lyrics \\
    \textbf{CFG Strategy}
      & LM-token drop
      & LM-token drop
      & LM-token drop
      & Full-drop
      & LM-token drop
      & Text-drop \\
    \midrule
    \multicolumn{7}{c}{\textbf{\textit{SongBench}}} \\
    \midrule
    Melody
      & 6.55 & 6.56 & 6.88 & \underline{6.89} & 6.88 & \textbf{6.96} \\
    Arrangement
      & 6.87 & 6.88 & 7.15 & \underline{7.17} & \underline{7.17} & \textbf{7.25} \\
    Musicality
      & 5.66 & 5.71 & 6.04 & \underline{6.05} & 6.04 & \textbf{6.17} \\
    Vocal
      & 6.80 & 6.88 & 7.21 & \underline{7.29} & 7.30 & \textbf{7.37} \\
    Instrumental
      & 6.96 & 6.95 & 7.08 & \underline{7.20} & 7.19 & \textbf{7.21} \\
    Mixing
      & 6.55 & 6.59 & 6.89 & \textbf{7.02} & \textbf{7.02} & \underline{7.01} \\
    Structure
      & 6.36 & 6.38 & 6.69 & \underline{6.75} & \underline{6.75} & \textbf{6.83} \\
    \midrule
    \multicolumn{7}{c}{\textbf{\textit{SongEval}}} \\
    \midrule
    Coherence
      & 4.28 & 4.31 & \underline{4.44} & 4.38 & 4.39 & \textbf{4.47} \\
    Musicality
      & 4.11 & 4.17 & \underline{4.28} & 4.25 & 4.25 & \textbf{4.34} \\
    Memorability
      & 4.18 & 4.24 & \underline{4.38} & 4.33 & 4.33 & \textbf{4.42} \\
    Clarity
      & 4.14 & 4.20 & \underline{4.33} & 4.29 & 4.29 & \textbf{4.37} \\
    Naturalness
      & 4.05 & 4.10 & \underline{4.22} & 4.18 & 4.18 & \textbf{4.26} \\
    \bottomrule
  \end{tabular}%
  }
\end{table*}

\subsubsection{Does the Band-Mode Refiner Improve Acoustic Reconstruction?}

We next isolate the acoustic decoder from LM generation and evaluate whether the Band-Mode Refiner improves reconstruction fidelity.
Spec-VAE and Spec-VAE\,+\,Refiner are evaluated on the full 546 tracks from the Song Describer Dataset~\citep{manco2023thesong}, together with four recent open-source audio VAE systems: $\varepsilon$ar-VAE~\citep{wang2025earVAE}, Stable Audio Open (SA-Open)~\citep{evans2024stableaudioopen}, Levo~2~\citep{lei2026levo2stablemelodious}, and SAME-L~\citep{parker2026samesemanticallyalignedmusicautoencoder}.
All systems are evaluated at their native operating conditions.
We report the baseline (\textbf{Spec-VAE}) and the full pipeline with Band-Mode Refiner (\textbf{Spec-VAE\,+\,Refiner}) separately to isolate the contribution of the refiner.

\paragraph{Metrics}
We measure reconstruction quality along three axes: spectral fidelity, mel-scale frequency response, and stereo image preservation. \textbf{(1) Spectral reconstruction:} We use SI-SDR (dB, $\uparrow$)~\citep{leroux2019sisdr}, STFT Distance ($\downarrow$), and STFT$_\text{log1p}$ ($\downarrow$) to measure time-domain fidelity, broadband spectral deviation, and low-amplitude spectral details. \textbf{(2) Mel reconstruction:} We use Mel Distance ($\downarrow$) and MEL$_\text{log1p}$ ($\downarrow$) to evaluate reconstruction quality on a perceptually spaced frequency axis. \textbf{(3) Stereo image quality:} 
We use CCPC (Cross-Channel Phase Coherence; $\in[0,1]$, $\uparrow$)~\citep{wang2025earVAE} and Spectral Pan Error ($\downarrow$). CCPC evaluates inter-channel phase coherence, while Spectral Pan Error measures the L1 distance in the frequency-wise stereo pan profile $\text{pan}(f) = |L(f)|^2 / (|L(f)|^2 + |R(f)|^2)$~\citep{avendano2002stereo}.

\begin{table*}[!tbh]
  \centering
  \small
  \scriptsize
  \renewcommand{\arraystretch}{0.84}
  \setlength{\tabcolsep}{4pt}
  \caption{
    Objective reconstruction quality on the Song Describer Dataset.
    All models are evaluated at their native operating conditions.
    Subscript values denote 95\% confidence intervals computed over the per-file metrics.
    Best results are highlighted in \textbf{bold}, and second-best results are \underline{underlined}.
  }
  \label{tab:decoder_objective}
  \resizebox{0.8\textwidth}{!}{
  \begin{tabular}{@{}lcccccc@{}}
    \toprule
    \textbf{Metric}
      & \makecell{\textbf{$\epsilon$ar-VAE}}
      & \makecell{\textbf{SA-Open}}
      & \makecell{\textbf{Levo 2}}
      & \makecell{\textbf{SAME-L}}
      & \makecell{\textbf{Spec-VAE}}
      & \makecell{\textbf{Spec-VAE}\\\textbf{+Refiner}} \\
    \midrule
    \multicolumn{7}{c}{\textbf{\textit{STFT-Scale Reconstruction}}} \\
    \midrule
    SI-SDR $\uparrow$
      & \underline{12.4}{\tiny\,$\pm$0.4}
      & 6.7{\tiny\,$\pm$0.3}
      & 8.1{\tiny\,$\pm$0.3}
      & \textbf{12.5}{\tiny\,$\pm$0.4}
      & 10.9{\tiny\,$\pm$0.4}
      & 11.3{\tiny\,$\pm$0.4} \\
    STFT Dist $\downarrow$
      & \underline{0.880}{\tiny\,$\pm$0.014}
      & 1.016{\tiny\,$\pm$0.016}
      & 0.971{\tiny\,$\pm$0.016}
      & 0.986{\tiny\,$\pm$0.015}
      & 0.916{\tiny\,$\pm$0.016}
      & \textbf{0.870}{\tiny\,$\pm$0.014} \\
    STFT$_\text{log1p}$ $\downarrow$
      & 0.079{\tiny\,$\pm$0.005}
      & 0.089{\tiny\,$\pm$0.005}
      & 0.086{\tiny\,$\pm$0.005}
      & 0.079{\tiny\,$\pm$0.005}
      & \underline{0.078}{\tiny\,$\pm$0.005}
      & \textbf{0.075}{\tiny\,$\pm$0.005} \\
    \midrule
    \multicolumn{7}{c}{\textbf{\textit{Mel-Scale Reconstruction}}} \\
    \midrule
    Mel Dist $\downarrow$
      & \underline{0.509}{\tiny\,$\pm$0.008}
      & 0.612{\tiny\,$\pm$0.007}
      & 0.599{\tiny\,$\pm$0.008}
      & 0.539{\tiny\,$\pm$0.010}
      & 0.572{\tiny\,$\pm$0.010}
      & \textbf{0.461}{\tiny\,$\pm$0.006} \\
    MEL$_\text{log1p}$ $\downarrow$
      & 0.095{\tiny\,$\pm$0.006}
      & 0.106{\tiny\,$\pm$0.006}
      & 0.103{\tiny\,$\pm$0.006}
      & 0.096{\tiny\,$\pm$0.006}
      & \underline{0.093}{\tiny\,$\pm$0.006}
      & \textbf{0.089}{\tiny\,$\pm$0.005} \\
    \midrule
    \multicolumn{7}{c}{\textbf{\textit{Stereo Image Quality}}} \\
    \midrule
    CCPC $\uparrow$
      & \textbf{0.973}{\tiny\,$\pm$0.003}
      & 0.933{\tiny\,$\pm$0.004}
      & 0.947{\tiny\,$\pm$0.004}
      & \underline{0.970}{\tiny\,$\pm$0.003}
      & 0.966{\tiny\,$\pm$0.003}
      & \textbf{0.973}{\tiny\,$\pm$0.002} \\
    Spectral Pan Err $\downarrow$
      & \underline{0.267}{\tiny\,$\pm$0.004}
      & 0.276{\tiny\,$\pm$0.004}
      & 0.273{\tiny\,$\pm$0.004}
      & 0.276{\tiny\,$\pm$0.004}
      & 0.268{\tiny\,$\pm$0.004}
      & \textbf{0.264}{\tiny\,$\pm$0.004} \\
    \bottomrule
  \end{tabular}}
\end{table*}

\paragraph{Results}
Table~\ref{tab:decoder_objective} summarizes the reconstruction results.
The full rendering decoder (\textbf{Spec-VAE\,+\,Refiner}) achieves the best STFT Distance (0.870), STFT$_\text{log1p}$ (0.075), Mel Distance (0.461), MEL$_\text{log1p}$ (0.089), and Spectral Pan Error (0.264), while matching the best CCPC score of 0.973.
Compared with Spec-VAE alone, the refiner improves all reported metrics, with the largest gain on Mel Distance (0.572 to 0.461).
These results support the band-specific correction strategy and show that the refiner improves perceptual frequency response and stereo spatial fidelity without sacrificing overall reconstruction quality.

\section{Conclusion}
\label{sec:conclusion}

In this report, we presented Qwen-Music, a unified large-scale music generation system for both text-to-music generation and reference-melody-based cover song generation.
Qwen-Music connects semantic-level composition with high-fidelity audio synthesis through three core components: \textbf{Qwen-Music-Tokenizer}, which represents music as compact 25 Hz single-codebook Music Semantic Tokens; \textbf{Qwen-Music-LLM}, which performs melody-aware autoregressive music modeling with a Melody-CoT mechanism; and \textbf{Qwen-Music-Render}, which performs generative stereo rendering from both Music Semantic Tokens and rewritten textual conditions.
Together, these components enable controllable full-song generation with coherent lyrics, stable musical structure, realistic vocals, and high-fidelity stereo waveforms.

To train this system at scale, we introduced a \textbf{quality-graded pre-training curriculum}, allowing Qwen-Music-LLM to first learn diverse musical patterns and then progressively improve generation quality.
We further adopted a \textbf{multi-stage post-training alignment} pipeline that combines supervised learning, iterative offline DPO, and online GSPO, improving musicality, instruction following, and controllability.
For Qwen-Music-Render, acoustic quality filtering provides reliable high-fidelity supervision for 48 kHz stereo rendering.

Extensive evaluations demonstrate that Qwen-Music achieves strong performance across subjective preference studies, genre-wise Bradley--Terry analysis, external leaderboard comparison, and objective metrics.
In text-to-music generation, Qwen-Music shows competitive or superior performance against leading proprietary systems in musicality, audio quality, controllability, and lyric intelligibility.
In cover song generation, it demonstrates strong reference-melody preservation and stylistic adaptation, while render-specific evaluations confirm the benefits of rewritten textual conditioning, text-drop CFG, and Band-Mode Refiner for high-fidelity stereo reconstruction.

Qwen-Music represents a step toward unified, scalable, and controllable music generation systems that bridge long-range musical semantics and waveform-level acoustic realization.
Future work will explore more flexible long-context musical structure modeling, finer-grained expressive control over singing and performance, and more efficient rendering architectures for high-quality music generation.

\clearpage
\bibliography{biblio}
\bibliographystyle{colm2024_conference}

\clearpage
\section{Authors}

\textbf{Core Contributors\footnotemark}
\begin{multicols}{6}
{\fontsize{9}{10}\selectfont
Jin Xu$^*$\\
Kangdi Wang\\
Ruibin Yuan\\
Shun Lei\\
Xiong Wang\\
Xize Cheng\\
Xueyao Zhang\\
Yang Zhang\\
Yiheng Chen\\
Yongqi Wang\\
Yue Wang\\
Zhifang Guo\\
Zhiyong Wu$^*$\\
Zihan Liu\\
Zijian Lin\\
}
\end{multicols}
\footnotetext{Alphabetical order. * denotes the corresponding author.}

\textbf{Contributors\footnotemark[\value{footnote}]}
\begin{multicols}{6}
{\fontsize{9}{10}\selectfont
Dake Guo\\
Hangrui Hu\\
Lei Xie\\
Linhan Ma\\
Wei Xue\\
Wenxiang Guo\\
Xinfa Zhu\\
Xipin Wei\\
Yangze Li\\
Yuanjun Lv\\
Yuxuan Wang\\
Yunfei Chu\\

}
\end{multicols}

\end{document}